\documentclass{iopart}

\usepackage{iopams}
\usepackage{graphicx}
\usepackage{cite}

\usepackage{perpage}
\MakePerPage{footnote}

\begin{document}
\title[]{Condensation transition in joint large deviations of linear statistics}

\author{J Szavits-Nossan$^1$, M R Evans$^1$ and S N Majumdar$^2$}

\address{$^1$ SUPA, School of Physics and Astronomy, University of Edinburgh, Mayfield Road, Edinburgh EH9 3JZ, United Kingdom}
\address{$^2$ Laboratoire de Physique Th\'{e}orique et Mod\`{e}les Statistiques, UMR 8626, Universit\'{e} Paris Sud 11 and CNRS, B\^{a}timent 100, Orsay F-91405, France}

\ead{jszavits@staffmail.ed.ac.uk}
\ead{mevans@staffmail.ed.ac.uk}
\ead{majumdar@lptms.u-psud.fr}

\begin{abstract}
Real space condensation is known to occur in stochastic models of mass transport in the regime in which the globally conserved mass density is greater than a critical value. It has been shown within models with factorised stationary states that the condensation can be understood in terms of sums of independent and identically distributed random variables: these  exhibit condensation when they are conditioned to a large deviation of their sum. It is well understood that the condensation, whereby one of the random variables contributes a finite fraction to the sum, occurs only if the underlying probability distribution (modulo exponential) is heavy-tailed, i.e. decaying slower than exponential. Here we study a similar phenomenon in which condensation is exhibited for non-heavy-tailed distributions, provided random variables are additionally conditioned on a large deviation of certain linear statistics. We provide a detailed theoretical analysis explaining the phenomenon, which is supported by Monte Carlo simulations (for the case where the additional constraint is the sample variance) and demonstrated in several physical systems. Our results suggest that the condensation is a generic phenomenon that pertains to both typical and rare events.
\end{abstract}


\submitto{\JPA}

\date{}

\maketitle

\section{Introduction}
\label{intro}

Real-space condensation occurs in models of mass transport when the globally conserved mass exceeds a critical value (see e.g. \cite{EvansHanney05} and \cite{SM_leshouches} for reviews). In this regime, the excess mass forms a condensate which is localised in space, for example at a single lattice site, and coexists with a background fluid in which the remaining mass is evenly distributed over the rest of the system. The condensation phenomenon occurs in many physical contexts such as coalescence in granular systems \cite{Torok05}, jamming in traffic \cite{OEC98,KMH05,LMZ04}, aggregation and absorption on surfaces \cite{MKB98,MKB00}, emulsification failure in polydisperse hard-sphere systems \cite{EMPT10}, etc. In these examples the `mass' corresponds to some conserved quantity such as the total number of particles, the total length of gaps between the cars or the volume fraction of hard spheres, respectively.

A particularly convenient case for analysing the condensation phenomenon is when the system has a factorised stationary state meaning that the stationary state probability of observing a configuration of masses $\{ m_i \}$, where $i=1\dots L$ is the site index, is
\begin{equation}
\label{fss}
P(m_1,\dots,m_L) = \frac{1}{Z_L(M)}\left[\prod_{i=1}^{L}f(m_i)\right]\delta\left(\sum_{j=1}^L m_j-M\right)\;.
\end{equation}
Here $Z_L(M)$ is the normalisation constant or nonequilibrium partition function; $f(m_i)$ is the single-site weight, and the delta function ensures that the total mass $M_L \equiv \sum_{i=1}^L m_i$ in the system is fixed to take value $M$. Thus the masses $m_i$ are uncorrelated except through this global constraint. 

A well-known model where the stationary state is given by (\ref{fss}) is the zero-range process in which $m_i$ is integer valued and the dynamics is such that one unit of mass moves from site $i$ to a neighbouring site with rate $u(m_i)$ which depends on the mass $m_i$ at the departure site. Then it may be shown that the single-site weight is given by $f(m_i) = \prod_{n=1}^{m_i} \frac{1}{u(n)}$, and by choosing different hopping rates $u(m)$ one obtains different forms for $f(m)$. Other models with  more complicated hopping rates also share a factorised stationary state \cite{WE12,EW14}  and exhibit condensation. However, it should be stressed that factorisation is not a pre-requisite for condensation.  More generally, models with continuous mass variables having factorised stationary states may be constructed \cite{EMZ04}. In any case, for both discrete and continuous masses, the condensation occurs when $f(m)$ decays slower than an exponential in $m$ \footnote{It is important to note that in (\ref{fss}) any exponential factor $a^m$ in $f(m)$ is immaterial as it results in a constant factor $a^M$ in the numerator which will be cancelled by a corresponding factor in $Z_L(M)$. Thus the condition for condensation can be extended to $f(m)$ having the form $f(m)=a^m \tilde{f}(m)$, provided $\tilde{f}(m)$ decays slower than an exponential.}. 

The condition for condensation of continuous masses\footnote{Similar condition can be written for integer-valued masses involving sums rather than integrals, see e.g. \cite{EvansHanney05}. For the purpose of establishing the connection with large deviation theory, we will focus here on continuous masses only.} may be written in terms of the mass density $\rho=M/L$ as
\begin{equation}
\rho > \rho_c = \frac{\int_{0}^{\infty} \rmd m\, m f(m)}{\int_{0}^{\infty}\rmd m f(m)}\;.
\label{cond}
\end{equation}
\noindent From (\ref{cond}) one sees that $\rho_c < \infty$ when
\begin{equation}
\int_{0}^{\infty} \rmd m\, m f(m) < \infty\;.
\label{ld}
\end{equation}
\noindent For example, a stretched exponential $f(m) \sim \exp(-a m^\alpha)$ with $\alpha <1$ or a power law $f(m) \sim A/m^b$ with $b>2$ fulfils (\ref{ld}). The ensuing condensation which occurs will be referred to as {\em standard} condensation. 
 
A clear signature of standard condensation is exhibited in the marginal distribution for the single-site mass which may be expressed as
\begin{equation}
\label{marginal_standard}
p(m)=f(m)\frac{Z_{L-1}(M-m)}{Z_L(M)}.
\end{equation}
\noindent In the condensed phase this distribution has a bump around the excess mass $M-\rho_cL$ as computed in\cite{MEZ05,EMZ06}. More rigorous work on standard condensation can be found in \cite{GrosskinskySchutzSpohn03,CG10,ArmendarizLoulakis11}.

As an aside we note that there are other variants of condensation such as strong condensation, which occurs for any mass density (i.e. $\rho_c=0$) when $f(m)$ increases with $m$ more quickly than exponentially \cite{JMP00,Jeon10}, and Bose-Einstein condensation which may occur when the hopping rate $u(m_i)$ depends on the site, leading to inhomogeneous weights $f_i(m_i)$ \cite{EvansHanney05}. Also a different form of condensation is exhibited in the inclusion process in the limit of certain rates tending to zero \cite{GRV11}.

Recently it has been appreciated that the condensation phenomenon, at least in the context of factorised stationary states, is related to large deviations of sums of random variables \cite{GrosskinskySchutzSpohn03,MEZ05,EMZ06,ArmendarizLoulakis11}. To see this we consider the normalisation in (\ref{fss})
\begin{equation}
\label{ZM}
Z_L(M)=\int_{0}^{\infty} \rmd m_1\dots \rmd m_L \left[\prod_{i=1}^{L}f(m_i)\right] \delta\left(\sum_{j=1}^{L}m_j-M\right)\;.
\end{equation}
\noindent  If we can normalise the single-site weight $f(m)$ so that $\int \rmd m f(m) =1$, then $P(m_1,\dots,m_L)$ in (\ref{fss}) is equivalent to the probability density of picking $L$ independent and identically distributed (iid) random variables with a common probability density $f(m)$, conditioned on the fixed value of their sample sum $M_L\equiv \sum_{i=1}^L m_i$. By fixing $M_L$ to take value $M$ that is far from the mean $\langle m\rangle L$, where (provided it exists) $\langle m \rangle$ is the average of $m$ with respect to a heavy-tailed $f(m)$, we can explore a regime where random variables are conditioned on the large deviation of $M_L$ and  employ results from large deviation theory.

Large deviation theory is concerned with the probability for events that are far away from the mean; the theory often predicts that the probability for such rare events is $\propto \exp(-LI(\Delta x))$, where $\Delta x$ is deviation from the mean; here, $L$ is called the speed of convergence and $I$ is the rate function. The mathematical theory of large deviations was first developed by Cram\'er in the 1930s for sums of iid random variables; it was later extended to correlated random variables resulting in the G\"{a}rtner-Ellis theorem which connects the cumulant generating function for the random variables to the rate function $I$ \cite{Ellis95}. 

The theory of large deviations has taken a prominent role in nonequilibrium statistical physics (for an overview, see \cite{Touchette09}). For example, the nonequivalence of microcanonical and canonical ensembles, characteristic of many systems with long-range interactions, is signalled by non-convexity of the rate function \cite{Touchette09}; in the context of condensation a similar nonequivalence of ensembles has been studied in \cite{GrosskinskySchutz08}. More generally, nonanalytic behaviour of $I$ signals a nonequilibrium phase transition \cite{BodineauDerrida05,GJLPDW09,Bertini10,Bunin12}. Recently, large deviation theory has been developed for stochastic systems of interacting particles such as the asymmetric simple exclusion process (ASEP), for the statistics of both the stationary \cite{DLS01} and time-dependent variables \cite{BodineauDerrida05}. Exact results from the ASEP were later also the main contributing factor in developing general theory for driven diffusive systems called the macroscopic fluctuation theory \cite{Bertini01,Bertini02,Bertini09} that uses large deviation theory extensively. 

A particularly striking result from large deviation theory concerns sums of iid random variables where $f(m)$ decays slower than an exponential \cite{Linnik61,Nagaev69}; such distributions are called heavy-tailed. In the large deviation regime, such that $M_L>\langle m\rangle L$, where $\langle m \rangle$ is the average of $m$ with respect to a heavy-tailed $f(m)$, one of the random variables typically takes a large value of $O(L)$ while the other $L-1$ take values of $O(1)$. The condition $M_L>\langle m\rangle L$ is precisely the condition for condensation (\ref{cond}) with critical mass density $\rho_c=\langle m\rangle$, provided we can relate single-site weight $f(m)$ to a probability density by a suitable normalisation. This distinctive feature of sums of heavy-tailed independent and identically distributed (iid) random variables is of main interest e.g. in financial modelling, in particular in risk theory, where condensation is a rare, but catastrophic event \cite{Embrechts97}.

In a previous communication \cite{SNEM14}, we reported that the condensation transition may be observed even when $f(m)$ is not heavy-tailed (henceforth termed light-tailed). The idea in \cite{SNEM14} was to introduce, in addition to fixing $M_L$, another global constraint by fixing the linear statistic
\begin{equation}
V_L\equiv \sum_i m_{i}^{1/p}\quad p\neq 1\;.
\label{Vdef}
\end{equation}
When this constraint is imposed, the light-tailed $f(m)$ may effectively become heavy-tailed allowing the standard, single-site, condensation transition to occur. A natural case to consider is $p=1/2$, from which  $V_L-M_{L}^{2}$ is the sample variance. We demonstrated how the condensation transition arises for the simplest case when $f(m)$ is an exponential distribution and for $p=1/2$; we also discussed briefly other $f(m)$ and general $p$. A particularly interesting case is when $f(m)$ is itself heavy-tailed, because the additional constraint in that case suppresses the condensation in the regime where it would have normally occurred with just $M_L$ fixed.

In this paper, we extend our work in \cite{SNEM14} in several ways. First, we provide alternative computations for general $p$ of the partition function 
in the presence of two constraints, $Z_L(M,V)$,  which is given by
\begin{eqnarray}
\label{ZMV}
Z_L(M,V)=&&\int_{0}^{\infty} {\rm d} m_1\dots {\rm d} m_L \left[\prod_{i=1}^{L}f(m_i)\right] \delta\left(\sum_{j=1}^{L}m_j-M\right)\nonumber\\
&&\times\delta\left(\sum_{k=1}^{L} m_{k}^{1/p}-V\right).
\end{eqnarray}
Second, we compute the marginal distribution $p(m)$, which is defined as
\begin{eqnarray}
p(m)&=&f(m)\int_{0}^{\infty}\rmd m_2\dots \rmd m_L \left[\prod_{i=2}^{L}f(m_i)\right]\delta\left(M_{L-1}-M+m\right)\nonumber\\
&&\times\delta\left(V_{L-1}-V+m^{1/p}\right)\nonumber\\
&=&f(m)\frac{Z_{L-1}(M-m,V-m^{1/p})}{Z_L(M,V)},
\label{marginal}
\end{eqnarray}
which should reveal a bump corresponding to the condensate in the condensed regime \cite{EMZ06}. We 
show that the condensate bump is shifted from the expected occupation number. We also provide numerical simulations in the case $p=1/2$ to support our calculations. Finally we discuss in detail some specific physical models that result in the constrained condensation phenomenon.

The remainder of the paper is organised as follows. Our main results are presented in Section \ref{main}, in particular the phase diagram in $M-V$ plane, which are then analysed in detail in Section \ref{analysis}. In Section \ref{MC} we present numerical simulations for the case $p=1/2$ that confirm our theoretical predictions. In Section \ref{examples}, we review several examples of condensation transition: jamming transition in exclusion process, condensation transition in polydisperse rods diffusing on a ring \cite{EMPT10} and phase transition in a random pure state of a large bipartite quantum system \cite{Nadal10,Nadal11}. We also mention a possibly related phenomenon of localised solutions (breathers) of the discrete nonlinear Schr\"odinger equation \cite{Rasmussen00,Johansson04,Rumpf,IPP14}.

\section{Main results}
\label{main}

We split our presentation of the main results in two cases, depending on whether the parameter $p$ in (\ref{Vdef}) takes values $0<p<1$ or $p>1$. Our main focus will be on the case $0<p<1$; the latter can be obtained from the former by making a suitable change of $f(m)$, as explained in Section \ref{case_p_1}. 

\subsection{Case $0<p<1$}
\label{case_0_p_1}

We consider $L$ real and non-negative iid random variables $m_i$, $i=1,\dots,L$ with common probability density $f(m)$, which are further conditioned on the fixed value of $M_L\equiv\sum_i m_i=M$ and $V_L\equiv\sum_i m_{i}^{1/p}=V$, where $0<p<1$. The probability density of finding a particular configuration $\{m_1,\dots,m_L\}$ for the given values of $M$ and $V$ reads
\begin{eqnarray}
\label{PMV}
P(m_1,\dots,m_L\vert M,V)=&&\frac{1}{Z_L(M,V)}\left[\prod_{i=1}^L f(m_i)\right]\delta\left(\sum_{j=1}^{L}m_j-M\right)\nonumber\\
&&\times\delta\left(\sum_{k=1}^{L}m_{k}^{1/p}-V\right),
\end{eqnarray}
\noindent where $Z_L(M,V)$ is the  normalisation constant defined in (\ref{ZMV}).
\noindent The probability density $P(m_i,\dots,m_L\vert M,V)$ in (\ref{PMV}) corresponds to a factorised steady state of a process whose dynamics conserve both $M_L$ and $V_L$. Alternatively, we can consider a standard mass-transfer model that only conserves $M_L$, and ask what is the probability distribution for $V_L$ given fixed $M_L=M$. In that case we are interested in the conditional probability density $P(V\vert M)$, which is given by
\begin{equation}
\label{PV_M}
P(V\vert M)=\frac{Z_L(M,V)}{Z_L(M)}\;,
\end{equation}
where $Z_L(M)$ is given by (\ref{ZM}).
Notice that both $Z_L(M,V)$ and $Z_L(M)$ have a simple probabilistic interpretations as discussed in the introduction. Namely, if we consider $m_i$ as iid random variables with common probability density $f(m)$, then $Z_L(M,V)$ is the joint probability density for random variables $M_L$ and $V_L$; similarly, $Z_L(M)$ is the probability density for $M_L$. 

Throughout the paper we assume that both $M$ and $V$ are large and proportional to $L$, so that we can write $M\equiv\mu L$ and $V\equiv\sigma L$. We also assume that $f(m)$ has finite moments
\begin{equation*}
\langle m\rangle  = \int_0^\infty \rmd m\,  m f(m)\quad \mbox{and}\quad \langle m^{1/p}\rangle = \int_0^\infty \rmd m\,  m^{1/p} f(m)\;.
\end{equation*}
\noindent Generally, we are interested in $\mu\neq\langle m\rangle$ and $\sigma\neq\langle m^{1/p}\rangle$, which means that $M_L$ and $V_L$ are far from their typical values $\langle m\rangle L$ and $\langle m^{1/p}\rangle L$, respectively.

Our main result concerns the phase diagram in the $\mu-\sigma$ plane, where phases differ in the behaviour of $Z_L(M,V)$ in the large-$L$ limit. Depending on the tail of $f(m)$, we find three qualitatively different cases, which are presented below. Notice also that in all cases we have the hard constraint
\begin{equation}
\sigma \geq \mu^{1/p}\quad\mbox{for}\quad 0 <p <1,
\label{hc}
\end{equation}
which follows from  Jensen's inequality:
\begin{equation*}
\phi(\langle m\rangle) \leq \langle \phi( m)\rangle \;,
\end{equation*}
\noindent when $\phi$ is a convex function.
 
%
{\bf Case (i).} Here for large $m$, $f(m)$ falls off as 
\begin{equation}
\label{case1}
f(m)\sim \rme^{-k m^{\gamma}}, \quad \gamma\geq 1/p, \quad k>0
\end{equation}

\begin{figure}[htb]
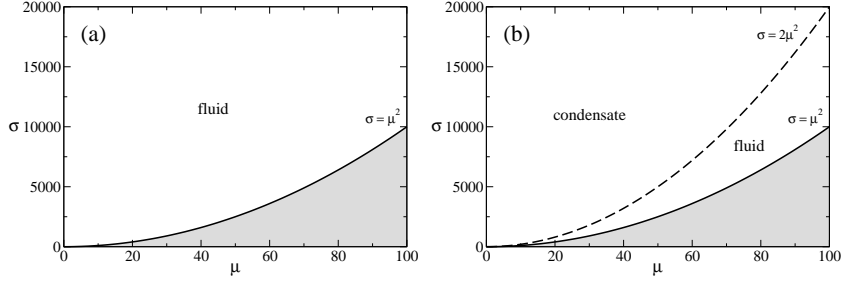

\centering\includegraphics[width=5.5cm]{fig1a.eps}
\centering\includegraphics[width=5.5cm]{fig1b.eps}
\caption{Phase diagram in $\mu-\sigma$ plane for $p=1/2$ and $f(m)$ that falls of: (a) {\bf Case (i)} faster than $\exp(-km^{1/p}))$, $k>0$, and (b) {\bf Case (ii)} exponentially or faster, but slower than in (a); the critical line in (b) can be calculated exactly for $f(m)=r\exp(-rm)$ and reads $\sigma_c(\mu)=2\mu^2$. Shaded area $\sigma<\mu^2$ is forbidden owing to Jensen's inequality.}
\label{fig1}
\end{figure}

\noindent In this case the large-$L$ behaviour of $Z_L(M,V)$ is given by
\begin{equation}
\label{ZMV_LD_1}
Z_L(M,V)\sim \rme^{-LJ(\mu,\sigma)}
\end{equation}
\noindent where the rate function $J(\mu,\sigma)$ is given by
\begin{equation}
\label{J_1}
J(\mu,\sigma)=-\lambda^*\sigma-s^*\mu-\textrm{ln}g(s^*,\lambda^*),
\end{equation}
\noindent and $s^*$ and $\lambda^*$ are defined implicitly via
\begin{eqnarray}
\mu&=&\frac{\int_{0}^{\infty}\rmd m\; mf(m)\rme^{-s^*m-\lambda^* m^{1/p}}}{\int_{0}^{\infty}\rmd m f(m)\rme^{-s^*m-\lambda^* m^{1/p}}}\label{saddle_mu}\\
\sigma&=&\frac{\int_{0}^{\infty}\rmd m\; m^{1/p} f(m)\rme^{-s^*m-\lambda^* m^{1/p}}}{\int_{0}^{\infty}\rmd m f(m)\rme^{-s^*m-\lambda^* m^{1/p}}}\label{saddle_sigma}.
\end{eqnarray} 
\noindent The system of equations (\ref{saddle_mu}) and (\ref{saddle_sigma}) can be solved for any $\mu$ and $\sigma$, and thus the corresponding phase diagram consists of just one phase (see e.g. phase diagram for $p=1/2$ displayed in Fig. \ref{fig1}(a)). 

We will show later that (\ref{ZMV_LD_1}) is a typical result from standard large deviation theory, and implies that large deviations of $M$ and $V$ are ``democratically'' spread across all random variables. This is more evident if we calculate the marginal distribution (\ref{marginal}), which we find takes the form
\begin{equation}
p(m)=f(m)\frac{\rme^{-s^*m-\lambda^* m^{1/p}}}{g(s^*,\lambda^*)},
\label{marginal1}
\end{equation}
\noindent where $g(s^*,\lambda^*)$ is the normalisation. Compared to the ``bare'' distribution $f(m)$, $p(m)$ acquires a factor $\exp(-s^* m-\lambda^* m^{1/p})$; also, there is no bump, i.e. all random variables contribute with small values to the sums $M_L$ and $V_L$. We will thus use the standard terminology ``fluid'' for this phase.

%
{\bf Case (ii).} A strikingly different situation happens if $f(m)$ falls as 
\begin{equation}
\label{case2}
f(m)\sim \rme^{-k m^{\gamma}}, \quad 1\leq\gamma< 1/p, \quad k>0
\end{equation}
\noindent The corresponding phase diagram, presented in Fig. \ref{fig1}(b) for the 
exponential distribution $f(m)=r\, \textrm{exp}(-rm)$ parametrised by $r$ and $p=1/2$, displays two phases separated by the critical line $\sigma_c(\mu)$: a fluid phase for $\mu^{1/p}\leq\sigma\leq \sigma_c(\mu)$ and a condensed phase for $\sigma>\sigma_c(\mu)$.

A fluid phase has the same large-$L$ behaviour as in case (i) (equation (\ref{ZMV_LD_1})), but now $\lambda^*$ must be non-negative in order for integrals in (\ref{saddle_mu}) and (\ref{saddle_sigma}) to converge; the critical line $\sigma_c(\mu)$ is defined by solving (\ref{saddle_mu}) and (\ref{saddle_sigma})  with $s^*\equiv r\geq 0$ and $\lambda^*=0$.

In the condensed phase, there is a condensate of size $\propto L^p$ residing on a single site, and the rest of the system is in the fluid phase. Behaviour of the $Z_L(M,V)$ for large $L$ is dominated by that of the $Z_L(M)$, i.e.
\begin{equation}
Z_L(M,V)\sim \rme^{-LI(\mu)},
\label{ZM_LD_2}
\end{equation}
\noindent where $I(\mu)$ is rate function of $Z_L(M)$ and is given by $I(\mu)=-\mu r-\textrm{ln}g(r,0)$ \cite{MEZ05,EMZ06}. We also find the following correction due to condensation, which has speed of convergence different from $L$,
\begin{equation}
\label{ZMV_LD_2}
\frac{Z_L(M,V)}{\rme^{-LI(\mu)}}\sim
\cases{\rme^{-k[L\sigma-L\sigma_c(\mu)]^{\gamma p}}, & $\gamma\neq 1$\cr
\rme^{-(k+s^*)[L\sigma-L\sigma_c(\mu)]^{ p}}, & $\gamma=1$\cr}.
\end{equation}
\noindent The above expression also applies to the conditional probability $P(V\vert M)$, since $P(V\vert M)=Z_L(M,V)/Z_L(M)\sim Z_L(M,V)/\exp(-LI(\mu))$. 

For the marginal distribution we have the following result,
\begin{equation}
\fl\quad p(m)\simeq f(m)\frac{\rme^{-rm}}{g(r,0)}\cdot
\cases{1, & $m\ll (L\sigma-L\sigma_c)^p$\cr
\frac{\exp[-\frac{(\mathbf{x}(m)-\mathbf{e})^T \mathbf{\Sigma}^{-1}(\mathbf{x}(m)-\mathbf{e})}{2(L-1)}]}{2\pi(L-1)\sqrt{\vert\Sigma}\vert Z_{L,r}(M,V)}, & $m\approx (L\sigma-L\sigma_c)^p$\cr},
\label{marginal2}
\end{equation}
\noindent where $Z_{L,r}(M,V)$ is a constant that depends on $r,M$ and $V$ and is given in (\ref{Z_r}). The bottom expression in (\ref{marginal2}), which corresponds to the condensate bump, is described by a bivariate Gaussian distribution, where the vector $\mathbf{x}(m)$, mean vector $\mathbf{e}$ and covariance matrix $\mathbf{\Sigma}$ are given in (\ref{multi_clt_2}) and (\ref{sigma}). Interestingly, as the non-diagonal elements of the covariance matrix $\mathbf{\Sigma}$ are non-zero, the position of the bump is generally shifted away from the expected occupation number of $(L\sigma-L\sigma_c)^p$. 
%
\begin{figure}[htb]
\centering\includegraphics[width=5.5cm]{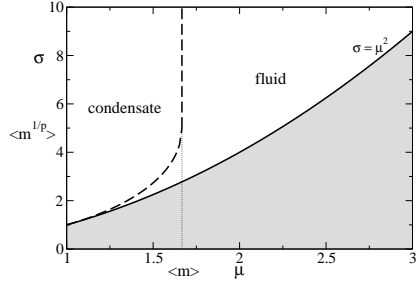}
\caption{Phase diagram in $\mu-\sigma$ plane for $p=1/2$ and $f(m)$ that falls off slower than exponential function ({\bf Case (iii)}; the critical line was calculated numerically for the Pareto distribution, 
$f(m)=(\gamma-1)/m^{\gamma}$ for $m>1$ and $\gamma=7/2$. Shaded area $\sigma<\mu^2$ is forbidden owing to Jensen's inequality.}
\label{fig2}
\end{figure}

{\bf Case (iii).} This is the case where $f(m)$ is heavy-tailed, i.e.
\begin{equation}
f(m)\quad \mbox{decays more slowly than} \exp(-km)\enskip \mbox{for any}\enskip k>0.
\label{case3}
%
\end{equation}
Recall that the standard condensation applies here when only $M_L$ is fixed provided $\mu>\langle m\rangle$, where $\langle m\rangle$ is the mean of $f(m)$ \cite{MEZ05,EMZ06}. However, this is no longer possible if $V_L$ is constrained to $V=L\sigma$, since the condensate would then imply $V_L\sim O(L^{1/p})$ and $1/p>1$. Instead, the condensation is suppressed for $\mu>\langle m\rangle$ and the critical line $\sigma_c(\mu)$ separating the fluid from the condensed phase extends thus from the point $(\mu,\sigma) =
(\langle m\rangle,\langle m^{1/p}\rangle)$ vertically as a straight line, as displayed in Fig. \ref{fig2} for 
Pareto distribution $f(m)=(\gamma-1)/m^\gamma$, $m>1$ and $\gamma=7/2$. The same phase diagram
was also recently found in Ref.~\cite{FZVM13} using a different method.

Similarly to the case (ii), behaviour of the $Z_L(M,V)$ and $P(V\vert M)$ for large $L$ is given by, respectively,
\begin{equation}
\label{ZMV_LD_3}
Z_L(M,V)\propto \rme^{-LI(\mu)}\rme^{-r[L\sigma-L\sigma_c(\mu)]^p},
\end{equation}

\begin{equation}
\label{PVM_LD_3}
P(V\vert M)\sim \frac{Z_L(M,V)}{\rme^{-LI(\mu)}}\sim\rme^{-r[L\sigma-L\sigma_c(\mu)]^p}.
\end{equation}

\noindent The marginal distribution takes the same form as in (\ref{marginal2}).

\subsection{Case $p>1$.}
\label{case_p_1}

This case is related to $0<p<1$ by suitable change of variables. Looking back at $Z_L(M,V)$ in (\ref{ZMV}) and making a change of variables $v_i=m_{i}^{q}$, where $q=1/p<1$, we obtain
\begin{eqnarray}
\label{ZMV_p_1}
Z_L(M,V)=&&\int_{0}^{\infty} {\rm d} v_1\dots {\rm d} v_L \left[ \prod_{i=1}^{L}f(v_{i}^{1/q})v_{i}^{1/q-1}/q\right]\delta\left(\sum_{j=1}^{L}v_j-V\right)\nonumber\\
&&\times\delta\left(\sum_k v_{k}^{1/q}-M\right).
\end{eqnarray}
\noindent We can thus use here all the results from Section \ref{case_0_p_1}, provided we substitute $V\leftrightarrow M$ (and thus $\mu\leftrightarrow\sigma$), $p\leftrightarrow q=1/p<1$ and $f(m)\rmd m\leftrightarrow [f(v^{1/q})v^{1/q-1}/q]\rmd v$. Obviously, the most significant difference here is that in the condensed regime one of the random variables $v_i\sim O(L^q)$, i.e. one of the original random variables $m_i\sim O(L)$, rather than of $ O(L^{p})$ as in the case $0<p<1$.

In the next Section we shall derive the main results for $0<p<1$ and
the compare them to the results from Monte Carlo simulations in Section 4.

\section{Detailed analysis of the phase diagram}
\label{analysis}

Our main focus is on the partition function $Z_L(M,V)$, equation (\ref{ZMV}), and its behaviour for large $L$. We shall first compute $Z_L(M,V)$ by computing its double Laplace transform $\tilde{Z}_L(s,\lambda)$, and then applying the saddle-point method to the inverse Laplace transform of $\tilde{Z}_L(s,\lambda)$, for large $L$. This procedure is standard in equilibrium statistical physics, where the grand canonical ensemble is used to remove hard constraints, in our case $M_L=M$ and $V_L=V$, and is then linked to the canonical ensemble via saddle-point equations. Alternatively (and more rigorously), we will use G\"{a}rtner-Ellis theorem to derive the rate function of $Z_L(M,V)$. Notably, none of the approaches will work in the condensed regime, where we will use large deviation theory for heavy-tailed sums instead.


\subsection{Computation of partition function through Laplace transforms}

For sums of iid random variables, the double Laplace transform of $Z_L(M,V)$ in equation (\ref{ZMV}) takes a factorised form
\begin{equation}
\tilde{Z}_L(s,\lambda)=\int_{0}^{M}dM\rme^{-sM}\int_{0}^{\infty}dV\rme^{-\lambda V}Z_L(M,V)=[g(s,\lambda)]^L,
\label{Z_inv}
\end{equation}
\noindent where $g(s,\lambda)$ in the last expression in (\ref{Z_inv}) is given by
\begin{equation}
g(s,\lambda)=\int_0^{\infty}\rmd m f(m)\rme^{-sm-\lambda m^{1/p}}.
\label{g}
\end{equation}
\noindent The partition function $Z_L(M,V)$ can be found by applying the saddle-point method to the inversion formula
\begin{equation}
Z_L(M,V)=\int_{c-i\infty}^{c+i\infty}\frac{ds}{2\pi i}\int_{d-i\infty}^{d+i\infty}\frac{d\lambda}{2\pi i}\rme^{Lh(s,\lambda)},
\label{ZMV_inv}
\end{equation}
\noindent where $c$ and $d$ are suitably chosen to be right of any singularities and $h(s,\lambda)$ is given by
\begin{equation}
h(s,\lambda)=s\mu+\lambda \sigma+\textrm{ln}g(s,\lambda).
\label{h}
\end{equation}
\noindent For large $L$, the largest contribution to the double integral in (\ref{ZMV_inv}) comes from the saddle point $s^*$, $\lambda^*$, implicitly defined via equations 
\begin{equation}
\mu=-\frac{\partial}{\partial s}\textrm{ln}g(s,\lambda),\quad \sigma=-\frac{\partial }{\partial \lambda}\textrm{ln}g(s,\lambda),
\label{saddle_2}
\end{equation}
\noindent which recover equations (\ref{saddle_mu}) and (\ref{saddle_sigma}). The question then arises as to when (\ref{saddle_2}) admit a solution. In this regard, it proves useful to introduce a function $z_q(s,\lambda)$ defined as
\begin{equation}
z_q(s,\lambda)=\frac{\int_{0}^{\infty}\rmd m\, m^{q} f(m)\rme^{-sm-\lambda m^{1/p}}}{g(s,\lambda)},
\label{z_q}
\end{equation}
\noindent where $q\geq 0$ is a parameter. Using (\ref{z_q}) we can rewrite saddle point equations (\ref{saddle_2}) as
\begin{equation}
\mu=z_1(s^*,\lambda^*),\quad \sigma=z_{1/p}(s^*,\lambda^*).
\label{saddle_3}
\end{equation}

We consider three cases with respect to the tail of $f(m)$:
\begin{eqnarray*}
&& \fl\quad\textrm{\bf Case (i):} \quad f(m)\sim \exp(-km^{\gamma}), \quad \gamma\geq 1/p, \quad k>0,\\
&& \fl\quad\textrm{\bf Case (ii):} \quad f(m)\sim \exp(-km^{\gamma}), \quad 1\leq\gamma<1/p, \quad k>0,\\
&& \fl\quad\textrm{\bf Case (iii):} \quad f(m) \;\textrm{decays more slowly than} \exp(-km)\enskip \mbox{for any} \enskip k>0
\end{eqnarray*}

\noindent In all three cases $z_1(s,\lambda)$ is a decreasing function of $s$ for fixed $\lambda$ and can take any value between $0$ and $\infty$ (see \ref{appendix_a}). Consequently, $\mu=z_1(s,\lambda)$ has an unique solution $s$ for any $\mu>0$; let us denote it with $s_\mu(\lambda)$. Furthermore, we show in \ref{appendix_a} that $z_{1/p}(s_\mu(\lambda),\lambda)$ is monotonically decreasing in $\lambda$ and decreases to $0$ in the limit $\lambda\rightarrow\infty$. The second equation in (\ref{saddle_3}) then admits a solution for any $\sigma\geq \mu^{1/p}$ ($\mu^{1/p}$ is the hard constraint due to Jensen's inequality) provided $z_{1/p}(s_\mu(\lambda),\lambda)$ is not bounded from above in the domain of allowed values of $\lambda$. The three cases above will differ precisely according to what that domain is, as follows.

In case (i), it is easy to see that $\lambda$ can take any value between $-\infty$ and $\infty$, and thus $\sigma=z_{1/p}(s_\mu(\lambda),\lambda)$ can be solved for any value of $\sigma>\mu^{1/p}$. In cases (ii) and (iii), $\lambda$ cannot be negative or otherwise the integrals in $z_{1/p}(s,\lambda)$ will diverge. In case (iii), where $f(m)$ decays more slowly than an exponential, we also require that if $\lambda=0$, $s$ must be positive. Thus, in cases (ii) and (iii) the maximum allowed values of $z_{1/p}(s_\mu(\lambda),\lambda)$ occur when $\lambda=0$. This will define a critical line
\begin{equation}
\sigma_c(\mu) = z_{1/p}(s_\mu(0),0)\;,
\label{sigmac}
\end{equation}
\noindent which separates the fluid phase from the condensed phase.


\subsection{Fluid phase $\mu^{1/p} \leq \sigma < \sigma_C$}
\label{fluid}

As noted in the introduction, in the fluid phase all random variables contribute with small values to the sums $M_L$ and $V_L$ and  the marginal distribution takes the form (\ref{marginal1}). In this Section we study the fluid phase in more detail using two alternative approaches: the saddle-point method and large deviation theory.

\subsubsection{Saddle-Point Method.}

To calculate $Z_L(M,V)$, we first expand $h(s,\lambda)$, defined in (\ref{h}), around $s=s^*$ and $\lambda=\lambda^*$
\begin{eqnarray}
h(s,\lambda)&=&h(s^{*},\lambda^{*})+\frac{1}{2}(s-s^*)^2[z_2(s^*,\lambda^*)-\mu^2]+\nonumber\\
&&+(s-s^*)(\lambda-\lambda^*)[z_{1+1/p}(s^*,\lambda^*)-\mu\sigma]\nonumber\\
&&+\frac{1}{2}(\lambda-\lambda^*)^2[z_{2/p}(s^*,\lambda^*)-\sigma^2]+\dots\;.
\label{h_exp}
\end{eqnarray}
\noindent Next, we insert (\ref{h_exp}) in (\ref{ZMV_inv}) with the choice of $c=s^*$ and $d=\lambda^*$ and make a change of variables $u=i(s-s^*)\sqrt{L}$ and $v=i(\lambda-\lambda^*)\sqrt{L}$ yielding
\begin{equation}
Z_L(M,V)=\frac{
me^{Lh(s^*,\lambda^*)}}{(2\pi)^2 L}\int_{-\infty}^{\infty}\int_{-\infty}^{\infty}\rmd u\rmd v \rme^{-\frac{1}{2}\mathbf{X}^{T}\mathbf{\Delta} \mathbf{X}+O(1/\sqrt{L})}.
\label{ZMV_saddle}
\end{equation}
\noindent Here the vector $\mathbf{X}$ and matrix $\mathbf{\Delta}$ are given by
\begin{equation*}
\fl\quad\vec{X}=\left(\begin{array}{c}u\\ v\end{array}\right),\quad 
\Delta=\left(\begin{array}{cc}
z_2(s^*,\lambda^*)-\mu^2 & z_{1+1/p}(s^*,\lambda^*)-\mu\sigma\\
z_{1+1/p}(s^*,\lambda^*)-\mu\sigma & z_{2/p}(s^*,\lambda^*)-\sigma^2\end{array}\right).
\end{equation*}
\noindent Ignoring terms of $\Or(1/\sqrt{L})$ in (\ref{ZMV_saddle}) and substituting $2\pi/\sqrt{\vert\mathbf{\Delta}\vert}$ for the double Gaussian integral gives the following  asymptotic expression for the partition function $Z_L(M,V)$ in the fluid phase 
\begin{equation}
Z_L(M,V)\simeq \frac{\rme^{Lh(s^*,\lambda^*)}}{2\pi L\sqrt{\vert\mathbf{\Delta}\vert}},\quad\mbox{for}\quad L\gg 1.
\end{equation}
\noindent This completes our derivation of the rate function $J(\mu,\sigma)=-s^*\mu-\lambda^*\sigma-\textrm{ln}g(s^*,\lambda^*)$ stated in (\ref{J_1}).

\subsubsection{Large-Deviation Approach.} In the fluid phase, we can also calculate the rate function $J(\mu,\sigma)$ in a more rigorous way using standard large deviation theory, in particular the G\"{a}rtner-Ellis theorem \cite{Ellis95,Touchette09}. Details of this calculation can be found in the \ref{appendix_b}.


\subsection{Critical line}
\label{critical}

As stated before, if $f(m)$ decays slower than $\exp(-km^{1/p})$ $\forall k>0$ ({\bf cases (ii)} and {\bf (iii)}), then the value $\lambda^*$ which solves (\ref{saddle_mu}) and (\ref{saddle_sigma}) cannot be negative. The limiting value of $z_{1/p}(s_\mu(\lambda),\lambda)$ when $\lambda\rightarrow 0$ gives the critical line (\ref{sigmac}) $\sigma_c(\mu)=z_{1/p}(s_{\mu}(0),0)$.

We also notice that the critical line contains the point $\mu= \langle m\rangle, \sigma = \langle m^{1/p}\rangle$ for which  $s_{\mu= \langle m\rangle}(0)=0$; this point splits the critical line in two segments: on the segment $\mu<\langle m\rangle$ the value of $s_\mu(0)$ is negative, and is positive for $\mu>\langle m\rangle$. However, the latter is not allowed if $f(m)$ is heavy-tailed as in case (iii); in that case the critical line extends vertically in a straight line (see Fig. \ref{fig2}). Interestingly, the second constraint $V_L=V$ has reduced the extent of the original condensed phase for the one constraint problem where only $M_L$ is constrained.


\subsection{Condensed phase}
\label{condensed}

In this Section we consider values of $\sigma>\sigma_c(\mu)$ and $f(m)$ that decays slower than $\exp(-k m^{1/p})$ $\forall k$, that is cases (ii) and (iii); in these cases we can no longer solve the equations (\ref{saddle_mu}) and (\ref{saddle_sigma}) and a different approach is needed. We show how to resolve this difficulty and how it implies the phenomenon of condensation.

Starting from the partition function $Z_L(M,V)$ defined in (\ref{ZMV}), we introduce a real number $r$ and insert $1=\exp(-rM)\exp(rM)$ in front of the integral in $Z_L(M,V)$
\begin{eqnarray*}
\fl\quad Z_L(M,V)&&=\rme^{-rM}\rme^{rM}\int_{0}^{\infty} {\rm d} m_1\dots {\rm d} m_L \left[\prod_{i=1}^{L}f(m_i)\right]\delta\left(\sum_{j=1}^{L}m_j-M\right)\\
\fl\quad &&\times\delta\left(\sum_{k=1}^L m_{k}^{1/p}-V\right)\\
\fl\quad &&=\rme^{rM}\int_{0}^{\infty} {\rm d} m_1\dots {\rm d} m_L \left[\prod_{i=1}^{L}f(m_i)\rme^{-rm_i}\right]\delta\left(\sum_{j=1}^{L}m_j-M\right)\\
\fl\quad &&\times\delta\left(\sum_{k=1}^L m_{k}^{1/p}-V\right),
\end{eqnarray*}
\noindent where in the last step we replaced $\exp(-rM)$ with $\exp(-r\sum_{j=1}^{L}m_j)$ as implied by the delta function. Next, we define a probability density $f_r(m)$,
\begin{displaymath}
f_r(m)=\frac{f(m)\rme^{-rm}}{g(r,0)},
\end{displaymath}
\noindent where $g(r,0)$ is the normalisation constant; $f_r(m)$ is sometimes called {\it twisted} or {\it exponentially tilted} distribution with respect to $f(m)$. We now make a change of variables $v_i=m_{i}^{1/p}$ and define a new probability density $w(v;r)$ parametrised by $r$ such that
\begin{equation}
f_r(m)\rmd m=pv^{p-1}f_r(v^{p})\rmd v\equiv w(v;r)\rmd v.
\label{w}
\end{equation}
\noindent Using $w(v;r)$, the partition function $Z_L(M,V)$ now reads
\begin{eqnarray}
Z_L(M,V)&&=\rme^{rM}[g(r,0)]^L\int_{0}^{\infty} {\rm d} v_1\dots {\rm d} v_L \left[ \prod_{i=1}^{L}w(v_i;r)\right]\nonumber \\
&&\times\delta\left(\sum_{j=1}^{L}v_{j}^p-M\right)\delta\left(\sum_{k=1}^L v_{k}-V\right). 
\label{ZMV_w}
\end{eqnarray}
\noindent Now, let us consider $L$ random variables $v_1,\dots,v_L$ that have a common probability density $w(v_i;r)$ and are conditioned on the value of their sum $\sum_{i=1}^{L}v_i=V$ (and thus dependent). As before, the probability density of finding a particular configuration $\{v_1,\dots,v_L\}$ can be written as
\begin{equation}
P_{L}^{w}(v_1,\dots,v_L;V)=\frac{1}{\Pi_L(V;r)}\left[\prod_{i=1}^{L}w(v_i;r)\right]\delta\left(\sum_{j=1}^{L}v_j-V\right),
\label{Pw}
\end{equation}
\noindent where $\Pi_L(V;r)$ is the normalisation constant given by
\begin{equation}
\label{Pi}
\Pi_L(V;r)=\int_{0}^{\infty}\rmd v_1\dots \rmd v_L\left[\prod_{i=1}^{L}w(v_i;r)\right]\delta\left(\sum_{j=1}^{L}v_j-V\right).
\end{equation}
\noindent Notice that the normalisation constant $\Pi_L(V;r)$ is itself a probability density for the sum $\sum_{i=1}^L v_i$, but where $v_1,\dots,v_L$ are {\it iid} random variables distributed by $w(v;r)$; we will use this fact later. Using (\ref{Pw}) and (\ref{Pi}), the expression for $Z_L(M,V)$ in (\ref{ZMV_w}) can be written as
\begin{equation}
Z_L(M,V)=\rme^{rM}[g(r,0)]^L \Pi_L(V;r)\left\langle\delta\left(\sum_{i=1}^{L}v_{i}^{p}-M\right)\right\rangle_{V},
\label{ZMV_r_w}
\end{equation}
\noindent where 
\begin{eqnarray}
\left\langle\delta\left(\sum_{i=1}^{L}v_{i}^{p}-M\right)\right\rangle_{V}&=&\frac{1}{\Pi_{L}(V;r)}\int_{0}^{\infty}\rmd v_1\dots \rmd v_L P_{L}^{w}(\{v_i\};V)\nonumber\\
&&\times \delta\left(\sum_{i=1}^{L}v_{i}^{p}-M\right).
\label{delta}
\end{eqnarray}

So far we have not specified the value for $r$. At this point we will choose $r$ for which the mean of $\sum_i m_i$, where $m_i$'s are iid random variables picked from the tilted distribution $f_r$, is equal to $\mu L$. In other words, we require that:
\begin{equation}
\label{r}
\mu=\int_{0}^{\infty}\rmd m\, m f_r(m)=z_1(r,0)\;.
\end{equation}
\noindent As discussed in Sections \ref{fluid} and \ref{critical}, we recall that (\ref{r}) can be solved for any $\mu>0$ provided $f(m)$ belongs to the case (ii) in (\ref{case2}). If $f(m)$ is heavy-tailed (case (iii) in (\ref{case3})), then (\ref{r}) can be solved for $0<\mu<\langle m\rangle$ (where the average is taken with respect to $f(m)$), which in fact encompasses the whole condensed phase (see Fig. \ref{fig2}).

Let us now go back to the definition for $w(v;r)$ in (\ref{w}). Using (\ref{r}), we can write
\begin{equation}
\langle v\rangle=\sigma_c(\mu)\quad \textrm{and}\quad \langle v^p\rangle=\mu,
\end{equation}
\noindent where the average is taken with respect to $w(v;r)$. This result tells us that in the condensed phase (where $\sigma>\sigma_c$), the sum $\sum_{i=1}^{L}v_i$ in (\ref{Pw}) and (\ref{Pi}) is conditioned to be a large deviation ($L\sigma$) compared to its mean ($L\sigma_c(\mu)$). However, in contrast to  the situation we had in the fluid phase, we can no longer apply the standard large deviation theory; this is due to the fact that $w(v;r)$ is heavy-tailed, and thus its moment-generating function, denoted by $\langle\exp(tv)\rangle$, diverges for all $t>0$. That $w(v;r)$ is indeed heavy-tailed can be seen by inspecting its tail. For $f(m)$ that decays as $\exp(-k m^{\gamma})$, where $\gamma< 1/p$  (case (ii) in (\ref{case2})), the right tail of $w(v;r)$ is given by
\begin{equation}
w(v;r)\sim \cases{
\rme^{-k v^{\gamma p}}, & Case (ii) in (\ref{case2}) and $1<\gamma<1/p$\cr
\rme^{-(k+r) v^p}, & Case (ii) in (\ref{case2}) and $\gamma=1$\cr}.
\label{w_2}
\end{equation}
\noindent Similarly, for heavy-tailed $f(m)$ (case (iii) in (\ref{case2})), the right tail of $w(v;r)$ is given by
\begin{equation}
w(v;r)\sim \rme^{-r v^{p}},\quad\textrm{Case (iii) in (\ref{case3})}.
\label{w_3}
\end{equation}
\noindent We see that in both cases $w(v;r)$ contains a stretched exponential tail that decays slower than an exponential. Since the sum $\sum_{i=1}^{L}v_i=L\sigma>L\langle v\rangle$ is conditioned to be a large deviation, the standard condensation follows in which, on average, $L-1$ random variables take the value $\langle v\rangle=\sigma_c(\mu)$ and one random variable takes the excess mass $L\sigma-L\sigma_c$. We can use this fact to estimate the unknown terms in (\ref{ZMV_r_w}), as follows. 

From \cite{Nagaev69} and \cite{EMZ06} we know that the tail of $\Pi_L(V;r)$ is determined by the tail of $w(L\sigma-L\sigma_c(\mu))$, that is, in the condensed regime
\begin{equation}
\Pi_L(V;r)\simeq  L w(L\sigma-L\sigma_c(\mu))\;.
\label{Picond}
\end{equation}
\noindent Expression (\ref{Picond}) comes from the $L$ ways to pick one of the random variables to be the condensate site multiplied by the weight of the condensate\footnote{One slight subtlety is that for heavy-tailed distributions that have stretched exponential tails $\exp(-v^\beta)$, (\ref{Picond}) strictly holds only for $0<\beta<1/2$; for $1/2<\beta<1$, $\Pi_L(V;r)$ has an additional factor due to finite contributions coming from the background fluid \cite{Nagaev69}; for a detailed discussion in the context of the zero-range process see \cite{ArmendarizGrosskinskyLoulakis13}.}. Therefore we deduce that the leading asymptotic behaviour of $\Pi_L(V;r)$ is
\begin{equation}
\fl\quad\Pi_{L}(V;r)\sim
\cases{\rme^{-k[L\sigma-L\sigma_c(\mu)]^{\gamma p}}, & {\bf Case (ii)} in (\ref{case2}) and $1<\gamma<1/p$\cr
\rme^{-(k+r)[L\sigma-L\sigma_c(\mu)]^{\gamma p}}, & {\bf Case (ii)} in (\ref{case2}) and $\gamma=1$\cr
\rme^{-r[L\sigma-L\sigma_c(\mu)]^p}, & {\bf Case (iii)} in (\ref{case3})\cr}\;.
\end{equation}

It now remains to estimate $\langle\delta(\sum_{i=1}^{L}v_{i}^{p}-M)\rangle_V$. We first note that $\langle\delta(\sum_{i=1}v_{i}^{p}-M)\rangle_V$ is the probability density for the sum $\sum_{i=1}^{L}v_{i}^{p}$ taking value of $M$, where random variables $v_1,\dots,v_L$ are distributed according to $P_{L}^{w}(v_1,\dots,v_L;V)$ in (\ref{Pw}). As we discussed above, the condition $\sum_{i=1}^{L}v_i=L\sigma>L\sigma_c(\mu)$ in (\ref{Pw}) leads to  condensation; the rest of $L-1$ random variables behave as if they were mutually independent and distributed with $w(v;r)$ \cite{GrosskinskySchutzSpohn03,ArmendarizLoulakis11}. In other words, the distribution of background fluid that co-exists with the condensate is given by the grand canonical distribution with maximal fugacity. Recalling that $\langle v^{p}\rangle=\mu$, typical values for the sum $\sum_{i=1}^{L}v_{i}^{p}$ are $(L-1)\mu+\Or(L^p)$, where the last term is due to the condensate. This heuristic argument implies that the sum $\sum_{i=1}^{L}v_{i}^{p}$ of random variables distributed with $P_{L}^{w}(v_1,\dots,v_L;V)$ fluctuates around $\mu L$, whereby the size of the fluctuations becomes increasingly small as $L\rightarrow\infty$. As a consequence, the hard constraint $\sum_{i=1}^{L}v_{i}^{p}=M$ will not affect the asymptotic behaviour  of $Z_L(M,V)$ for large $L$ and can safely be ignored.

Using the fact that $Z_L(M) \sim \exp(-rM)[g(r,0)]^L=\exp[-LI(\mu)]$, we can finally write  
\begin{equation}
\fl\quad\frac{Z_L(M,V)}{Z_L(M)}\sim
\cases{\rme^{-k[L\sigma-L\sigma_c(\mu)]^{\gamma p}}, & Case (ii) in (\ref{case2}) and $\gamma\neq 1$\cr
\rme^{-(k+r)[L\sigma-L\sigma_c(\mu)]^{ p}}, & Case (ii) in (\ref{case2}) and $\gamma=1$\cr
\rme^{-r[L\sigma-L\sigma_c(\mu)]^p}, & Case (iii) in (\ref{case3})\cr},
\label{Zcond}
\end{equation}
\noindent which recovers results stated in (\ref{ZMV_LD_2}), (\ref{ZMV_LD_3}) and (\ref{PVM_LD_3}). 


\subsection{Marginal probability}

The most striking way to demonstrate the condensation is to calculate the marginal distribution $p(m)$, which is defined in (\ref{marginal}).
As noted in the introduction a bump in $p(m)$ corresponding to the condensate appears in the condensed phase.

In the fluid phase, $Z_L(M,V)\propto \exp(s^*M+\lambda^*V)$ which gives
\begin{equation}
p(m)\approx f(m)\frac{\exp^{-s^*m-\lambda^*m}}{g(s^*,\lambda^*)}.
\label{pmfluid}
\end{equation}
\noindent where $s^*$ and $\lambda^*$ are the solutions of (\ref{saddle_mu}) and (\ref{saddle_sigma}).

In the condensed phase, we can use the idea from Section \ref{condensed}, where we introduced the tilted distribution $f_r(m)=f(m)\exp(-rm)/g(r,0)$. This leads to
\begin{equation}
\label{marginal_2}
\fl\quad p(m)=f_r(m)\frac{\Pi_{L-1}(V-m^{1/p};r)\left\langle\delta\left(\sum_{i=1}^{L-1}v_{i}^{p}-M+m\right)\right\rangle_{V-m^{1/p}}}{\Pi_{L}(V;r)\left\langle\delta\left(\sum_{i=1}^{L}v_{i}^{p}-M\right)\right\rangle_V}.
\end{equation}

For $m=\Or(1)$, we expect the ratio in (\ref{marginal_2}) to be close to $1$ yielding
\begin{equation}
p(m)\approx f(m)\frac{\rme^{-rm}}{g(r,0)},\quad m\ll (V-V_c)^p.
\label{marginal_cond_low}
\end{equation}

For large $m$, we would expect to find a bump peaked at $m=(V-V_c)^{p}$; instead, we find that the centre of the bump is slightly shifted. To understand this shift, it  proves useful to rewrite $p(m)$ as 
\begin{equation}
\fl\quad p(m)=f_r(m)\frac{\int_{0}^{\infty}\prod_{i=1}^{L-1}\rmd m_i f_{r}(m_i)\delta(M_{L-1}-M+m)\delta(V_{L-1}-V+m^{1/p})}{Z_{L,r}(M,V)}\;,
\label{marginal_joint}
\end{equation}
\noindent where $Z_{L,r}(M,V)$ is given by
\begin{equation}
Z_{L,r}(M,V)=\int_{0}^{\infty}\prod_{i=1}^{L}\rmd m_i f_{r}(m_i)\delta(M_{L}-M)\delta(V_{L}-V).
\label{Z_r}
\end{equation}

\noindent Here the integral in the numerator in (\ref{marginal_joint}) is {\it joint} probability for the sums $M_{L-1}$ and $V_{L-1}$ of {\it iid} random variables distributed with $f_r$. 

For $m$ close to $(L\sigma-L\sigma_c)^p$, $M_{L-1}=M-m$ is only of $\Or(L^p)$ away from its mean $(L-1)\mu$; similarly, $V_{L-1}=V-m^{1/p}\approx L\sigma_c$ is very close to its mean $(L-1)\sigma_c$. For these values of $M_{L-1}$ and $V_{L-1}$ there is no condensation, and we assume that random variables $m_i$ can be thus treated as independent. We now use this approximation as an heuristic to determine the bump in $p(m)$ by approximating the joint probability density for $M_{L-1}$ and $V_{L-1}$ by a bivariate Gaussian distribution, according to the central limit theorem (see \ref{appendix_b} for details)
\begin{eqnarray}
&&\int_{0}^{\infty}\prod_{i=1}^{L-1}\rmd m_i f_{r}(m_i)\delta(M_{L-1}-M+m)\delta(V_{L-1}-V+m^{1/p})\nonumber\\
\fl &&\approx\quad\frac{1}{2\pi (L-1)\sqrt{\vert\mathbf{\Sigma}\vert}}\exp\left[-\frac{(\mathbf{x}(m)-\mathbf{e})^T \mathbf{\Sigma}^{-1}(\mathbf{x}(m)-\mathbf{e})}{2(L-1)}\right],
\label{multi_clt}
\end{eqnarray}
\noindent where $\mathbf{x}(m)$, $\mathbf{e}$ and $\Sigma$ are given by, respectively,
\begin{equation}
\mathbf{x}(m)=\left(\begin{array}{c} M-m \\ V-m^{1/p} \end{array}\right),\quad \mathbf{e}=\left(\begin{array}{c} (L-1)\mu \\ (L-1)\sigma_c \end{array}\right), \quad \textrm{and}
\label{multi_clt_2}
\end{equation}
\begin{equation}
\mathbf{\Sigma}=\left(\begin{array}{cc}
z_2(r,0)-\mu^2 &  z_{1+1/p}(r,0)-\mu\sigma_c \\ 
z_{1+1/p}(r,0)-\mu\sigma_c & z_{2/p}(r,0)-\sigma_{c}^{2}\end{array}\right).
\label{sigma}
\end{equation}
\noindent Inserting (\ref{multi_clt}) in (\ref{marginal_joint}) recovers the expression for $p(m)$ stated in (\ref{marginal2}). Note that the above argument is not expected to hold when $1/2<p<1$, as in that case the value $M-m$ of the sum $M_{L-1}$ falls out of the zone where central limit theorem normally applies. However, as long as $p<1$, one can in principle get higher-order corrections using e.g. multivariate Edgeworth expansions.

Finally, to calculate the shift from the naively expected value of $m_{\mathrm{cond}}=(V-V_c)^p$, we look for the maximum of $p(m)$, which solves the following equation:
\begin{equation}
\frac{f'(m)}{f(m)}-r-\frac{1}{2(L-1)}\frac{d}{dm}\left[(\mathbf{x}(m)-\mathbf{e})^T \mathbf{\Sigma}^{-1}(\mathbf{x}(m)-\mathbf{e})\right]=0.
\label{shift}
\end{equation}
\noindent By solving this equation for $L\gg 1$, we find the location of the bump $m_{\mathrm{cond}}$
\begin{equation}
m_{\mathrm{cond}}=L^{p}(\sigma-\sigma_c)^p+\epsilon,
\end{equation}
\noindent where $\epsilon$ is given to leading order by
\begin{equation}
\fl\epsilon\simeq
\cases{
L^{2p-1}(\sigma-\sigma_c)^{2p-1}\left(\frac{p\Sigma_{12}}{\Sigma_{11}}-\frac{p^2 (r+k)\vert\mathbf{\Sigma}\vert}{\Sigma_{11}(\sigma-\sigma_c)}\right), & {\bf case (ii)}, $\gamma=1$\cr
-L^{\gamma p-1+p}(\sigma-\sigma_c)^{\gamma p-2+p}\frac{k \gamma p^2\vert\mathbf{\Sigma}\vert}{\Sigma_{11}}, & {\bf case (ii)}, $1<\gamma<1/p$\cr
L^{2p-1}(\sigma-\sigma_c)^{2p-1}\left(\frac{p\Sigma_{12}}{\Sigma_{11}}-\frac{p^2 r\vert\mathbf{\Sigma}\vert}{\Sigma_{11}(\sigma-\sigma_c)}\right), & {\bf case (iii)}\cr}
\label{epsilon}
\end{equation}
We note that $\epsilon$ is subdominant as $\gamma<1/p$ and $p<1$, but it may still diverge with $L$ for $p$ sufficiently large. This means that a shift 
in the condensate bump position will be observed even for large $L$. 

In the next section we will demonstrate the condensation transition by constructing a Markov process that generates random variables under two constraints. In this way we can have condensation as a typical event, rather than search for it in rare fluctuations of $V_L$. Our simulations are restricted to $p=1/2$; other integer values of $1/p$ are possible in theory, but are difficult to implement.

\section{Monte Carlo simulations}
\label{MC}

In this Section we will conduct Monte Carlo simulations to test our theoretical predictions for the condensation under two constraints. Generating random numbers $m_i$  that satisfy both $M_L=M$ and $V_L=V$ is generally a difficult problem. To this end, we will consider only the case $p=1/2$, which allows us to construct a stochastic process in which both $M_L$ and $V_L$ are fixed \cite{IPP14}. 

\subsection{Algorithm for the case $p=1/2$}

We use an algorithm introduced in \cite{IPP14} to sample the distribution (\ref{PMV}). We consider a chain of $L$ sites with periodic boundary conditions, where each site carries a mass $m_i\geq 0$, $i=1,\dots,L$. The continuous time dynamics is approximated by the following random sequential update rule. In each time increment $t \to t +\Delta t$, we choose a site $i$ at random and look at the triplet $\{m_{i-1},m_i,m_{i+1}\}$. Let for this particular timestep
\numparts
\begin{eqnarray}
&& m_{i-1}(t)+m_{i}(t)+m_{i+1}(t)\equiv\rho(t)\label{rho_t}\\
&& m_{i-1}^{2}(t)+m_{i}^{2}(t)+m_{i+1}^{2}(t)\equiv\omega(t).\label{omega_t}
\end{eqnarray}
\endnumparts

\noindent The idea is then to replace $\{m_{i-1}, m_i, m_{i+1}\}$ with randomly chosen values $\{m'_{i-1}, m'_i, m'_{i+1}\}$ which also satisfy $m'_{i-1}+m'_{i}+m'_{i+1}=\rho(t)$ and ${m'}_{i-1}^{2}+{m'}_{i}^{2}+{m'}_{i+1}^{2}=\omega(t)$. Then we set $\{m_{i-1}(t+\Delta t), m_i(t + \Delta t), m_{i+1}(t + \Delta t)\}=\{m'_{i-1}, m'_i, m'_{i+1}\}$. When the process reaches the stationary state we are  able to construct empirically the marginal distribution $p(m)$ from the values $m_i$.

\begin{figure}[htb]
\centering\includegraphics[width=4cm]{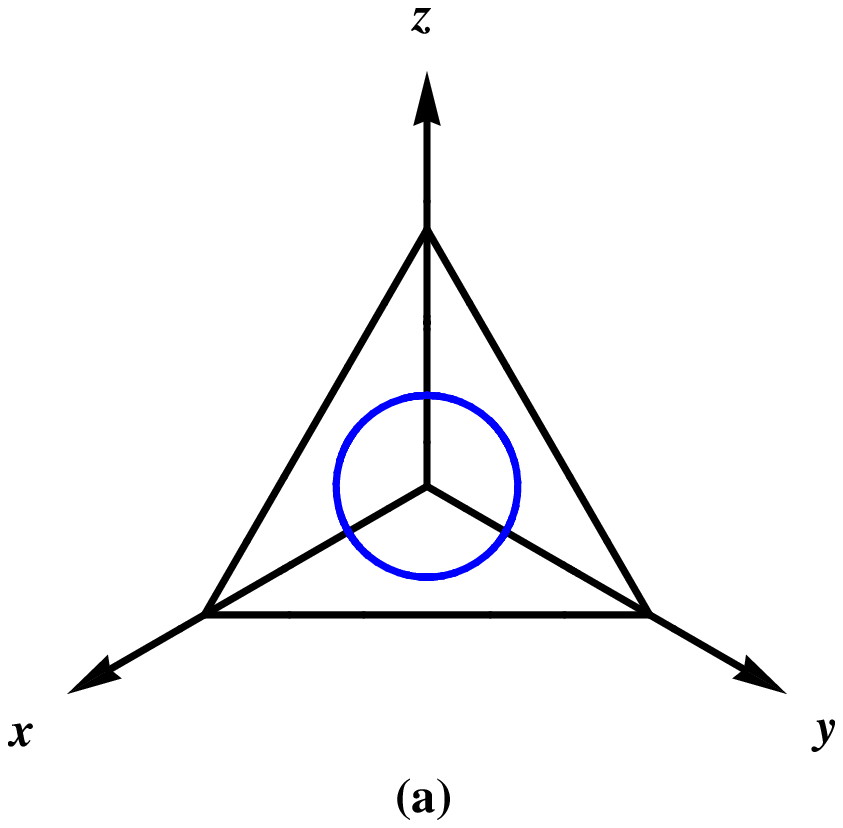}
\hspace{1cm}
\centering\includegraphics[width=4cm]{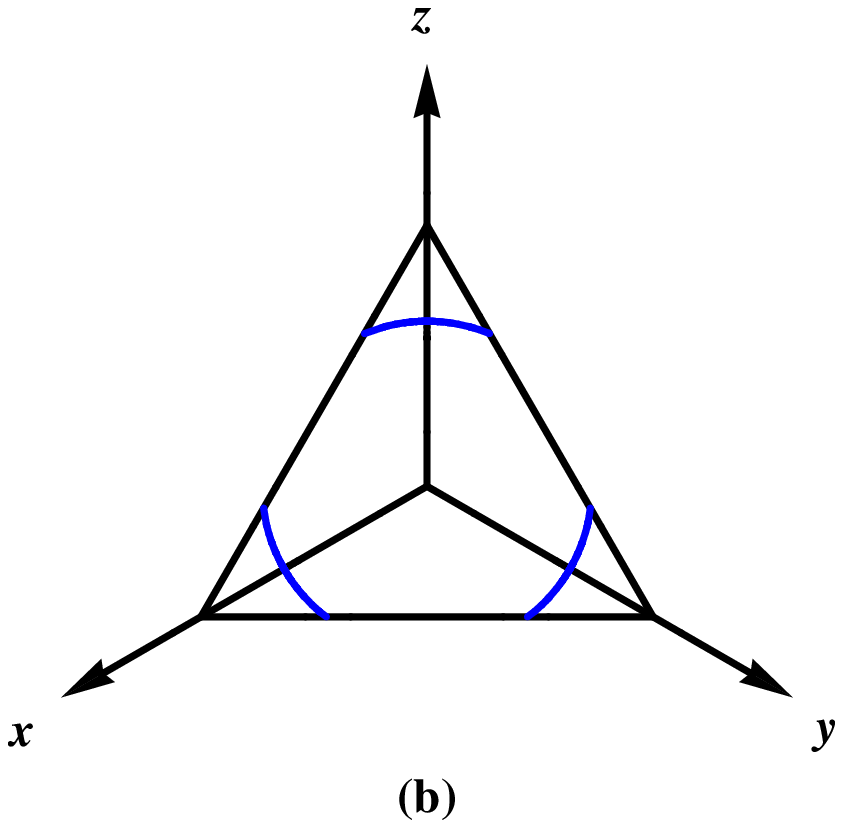}
\caption{Intersection (blue line) of the plane $x+y+z=\rho$ and the sphere $x^2+y^2+z^2=\omega$ for $\rho^2/3\leq\omega\leq\rho^2$. The condition $x\geq 0,y\geq 0$ and $z\geq 0$ yields either a circle for $\rho^2/3<\omega\leq\rho^2/2$ (left) or disjoint arcs for $\rho^2/2\leq\omega\leq\rho^2$ (right).}
\label{fig3}
\end{figure}

In order to choose values of $\{m'_{i-1},m'_i,m'_{i+1}\}$, which we denote $ \{x,y,z\}$, that satisfy the constraints (\ref{rho_t}) and (\ref{omega_t}), we are looking for the intersection of the plane $x+y+z=\rho$ and the sphere $x^2+y^2+z^2=\omega$, under condition that $x,y,z\geq 0$. Without the latter, the result is a circle of radius $\sqrt{\omega-\rho^2/3}$ parametrised by an angle $\theta$
\begin{eqnarray}
x(\theta;\rho,\omega)&=&\frac{\rho}{3}-\sqrt{\frac{2}{3}\left(\omega-\frac{\rho^2}{3}\right)}\sin\left(\theta+\frac{\pi}{3}\right)\label{x}\\
y(\theta;\rho,\omega)&=&\frac{\rho}{3}-\sqrt{\frac{2}{3}\left(\omega-\frac{\rho^2}{3}\right)}\sin\left(\theta-\frac{\pi}{3}\right)\label{y}\\
z(\theta;\rho,\omega)&=&\frac{\rho}{3}+\sqrt{\frac{2}{3}\left(\omega-\frac{\rho^2}{3}\right)}\sin\theta.\label{z}
\end{eqnarray} 
The condition $x,y,z\geq 0$ yields two possible ranges for $\theta$ according to the values of $\rho$ and $\omega$: $\theta\in [0,2\pi]$ for $\rho^2/3<\omega\leq\rho^2/2$ (figure \ref{fig3}(a)) and $[\pi/6+\alpha,5\pi/6-\alpha]$, $[5\pi/6+\alpha,9\pi/6-\alpha]$ and $[9\pi/6+\alpha,13\pi/6-\alpha]$ for $\rho^2/2<\omega<\rho^2$ (figure \ref{fig3}(b)), where $\alpha=\textrm{arccos}(\rho/\sqrt{6\omega-2\rho^2})$.
%

In each step, we  choose $\theta'$ uniformly on the allowed domain and set $m'_{i-1}=x(\theta';\rho,\omega)$, $m'_i=y(\theta';\rho,\omega)$ and $m'_{i+1}=z(\theta';\rho,\omega)$\footnote{Here, in the case of disjoint allowed arcs we  choose $\theta'$ uniformly from all three disjoint arcs, whereas in the algorithm used in \cite{IPP14} $\theta'$ was restricted to the same arc as $\theta$}. This algorithm generates a stationary state of the form (\ref{PMV}), with the single site weight $f(m)=1$ \footnote{Here $f(m)=1$ is no longer a probability density, as it cannot be normalised on $[0,\infty\rangle$. However, due to fixed $M_L$ and $V_L$, this case is closely related to having an exponential distribution or a Gaussian (or a mix of both).},
\begin{displaymath}
P(m_1,\dots,m_L)=\frac{\delta\left(\sum_i m_i-M\right)\delta\left(\sum_i m_{i}^{2}-V\right)}{Z_L(M,V)},
\end{displaymath}
\noindent where the normalisation constant $Z_L(M,V)$ is given by
\begin{equation}
\fl\quad Z_L(M,V)=\int_{0}^{\infty}\rmd m_1\dots \rmd m_L\delta\left(\sum_i m_i-M\right)\delta\left(\sum_j m_{j}^{2}-V\right).
\end{equation}
\noindent In this particular case $\sigma_c(\mu)$ can be found exactly and reads $\sigma_c(\mu)=2\mu^2$ \cite{Chatterjee10,SNEM14}.

The algorithm can be further generalised to generate stationary distributions of the form (\ref{PMV}) with $f(m) \neq 1$ by choosing transition rates $W(\{m_i\}\rightarrow\{m'_{i}\})$ that satisfy the detailed balance condition:
\begin{equation}
\label{rates}
W(\{m_i\}\rightarrow\{m'_{i}\})=\frac{f(m'_{i-1})f(m'_i)f(m'_{i+1})}{f(m_{i-1})f(m_i)f(m_{i+1})} W(\{m'_i\}\rightarrow\{m_{i}\}).
\end{equation}
To implement this, we can use the standard Metropolis algorithm where the candidate update $\{m_i\}\rightarrow\{m'_{i}\}$ is accepted with the following probabilities
\begin{eqnarray}
\{m_i\}\rightarrow\{m'_{i}\}\quad\textrm{with probability}\quad\cases{
1, & if $\alpha>1$\cr
\alpha, & if $\alpha<1$\cr},
\end{eqnarray}
\noindent where $\alpha$ is given by
\begin{equation*}
\alpha=\frac{f(m'_{i-1})f(m'_i)f(m'_{i+1})}{f(m_{i-1})f(m_i)f(m_{i+1})}.
\end{equation*}

\subsection{Numerical results}

\begin{figure}[htb]
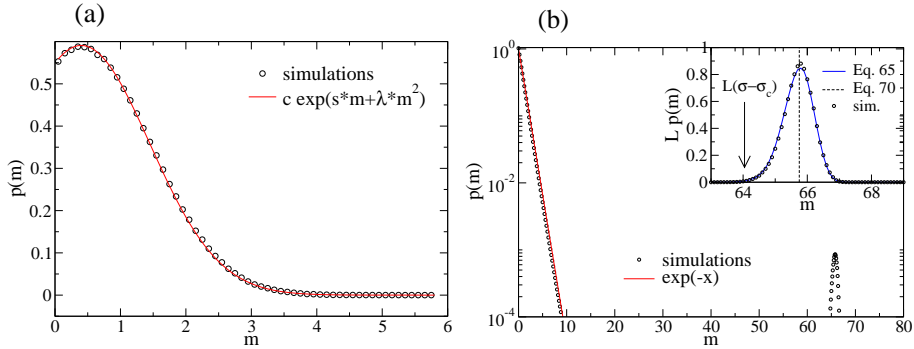

\centering\includegraphics[width=6cm]{fig4a.eps}
\centering\includegraphics[width=6cm]{fig4b.eps}
\caption{Marginal distribution for (a) $\sigma=3/2<\sigma_c(\mu)=2$ (fluid phase) and (b) $\sigma=6>\sigma_c(\mu)=2$ (condensed phase), obtained from $ 10^8$ random numbers generated using the algorithm described in the text, for $L=1024$ and $\mu=1$. In (a), $p(m)$ is compared to $c \exp(-s^*m-\lambda^*m^2)$ (full line), where $s^*(\lambda^*)\approx -0.36503$ and $\lambda^*\approx 0.45501$ were calculated by solving (\ref{saddle_mu}) and (\ref{saddle_sigma}) numerically and $c\approx 0.54921$ was chosen to best fit the data. In (b), $p(m)$ is compared to $\exp(-r m)$ (full line) for $m\ll\sqrt{L(\sigma-\sigma_c)}$, where $r=1/\mu=1$ solves (\ref{r}). In the inset: marginal distribution for $m\approx\sqrt{V-V_c}$, compared to the theoretical prediction in (\ref{marginal_2}) obtained using (\ref{multi_clt}) (dashed line); the arrow is pointing to the naive prediction $\sqrt{L\sigma-L\sigma_c}=64$ and the dashed line is the prediction including the shift $\epsilon=1.75$ given by (\ref{epsilon}) and (\ref{shift_f1}).}
\label{fig4}
\end{figure}

Monte Carlo simulations were conducted for $\sigma=3/2<\sigma_c=2$ (fluid phase, Figure \ref{fig4}(a)) and $\sigma=6>\sigma_c=2$ (condensed phase, Figure \ref{fig4}(b)). In total $5\cdot 10^8$ sets of numbers $\{m_1,\dots,m_L\}$ were generated with fixed $M_L=\mu L$ and $V_L=\sigma L$ for $L=1024$ and $\mu=1$, using the algorithm described above. At the beginning, we assigned mass $m_+$ to $\phi L$ randomly chosen sites and the rest were assigned to $m_-$, where $\phi$ can be any positive number less than or equal to $\mu^2/\sigma$; here $m_-$ and $m_+$ were chosen to satisfy the two constraints yielding
\begin{equation}
m_-=\mu-\sqrt{\frac{\phi}{1-\phi}(\sigma-\mu^2)},\quad m_+=\mu+\sqrt{\frac{1-\phi}{\phi}(\sigma-\mu^2)}.
\end{equation}
\noindent The marginal distribution was calculated by counting the number of particles in bins of sizes $0.1$ every $5$ Monte Carlo sweeps to reduce correlations (one Monte Carlo sweep comprises $L$ updates). In total $10^8$ sets $\{m_1,\dots,m_L\}$ were used to calculate the marginal distribution, presented in figures \ref{fig4}(a) and \ref{fig4}(b).

From Fig. \ref{fig4}(a), we see that $p(m)$ is indeed given by (\ref{pmfluid}) in the fluid phase, where $s^*$ and $\lambda^*$ were obtained by solving (\ref{saddle_mu}) and (\ref{saddle_sigma}) numerically. In the condensed phase (Fig. \ref{fig4}(b)), $p(m)$ is in excellent agreement with (\ref{marginal}) for $m\ll \sqrt{L(\sigma-\sigma_c)}$; for $m$ close to $\sqrt{L\sigma-L\sigma_c}$, a clear bump corresponding to the condensate appears with the centre slightly shifted to the right by an amount that is in good agreement with (\ref{epsilon}) in the case (ii) for $\gamma=1$ and $k=0$ (which corresponds to $f(m)=1$)
\begin{equation}
\epsilon\simeq 2-\frac{1}{\sigma-\sigma_c}.
\label{shift_f1}
\end{equation}
Moreover, the shape of the bump is in very good agreement with our prediction in (\ref{multi_clt}), as can be seen in the inset of Fig. \ref{fig4}(b); no fitting parameters were used in Fig. \ref{fig4}, except for the normalisation constant $Z_{L}(M,V)$  which was calculated numerically by normalizing mass distribution to $1$.

\section{Examples}
\label{examples}


In this Section we discuss several examples of physical systems with constraints. Apart from demonstrating the condensation phenomenon, the examples imply a general mechanism by which condensation can occur either as a typical event (when both hard constraints are present) or alternatively, as a rare event (when there is only one hard constraint).

\subsection{Spontaneous jamming in the exclusion process}

Our first example is the asymmetric simple exclusion process (ASEP), a one-dimensional system of particles interacting via exclusion principle (for a review see e.g.\cite{BlytheEvans07}). In the ASEP, $N=\rho L$ particles hop on an one-dimensional lattice of $L$ sites such that at any moment no site holds more than one particle. Of several possible rules to move the particles, here we will consider the continuous time dynamics in which each particle attempts hops with rate one per unit time, either one site to the left with probability $q$ or one site to the right with probability $p$. For simplicity, we consider the totally asymmetric case $q=0$ with $p=1$. We also assume periodic boundary conditions, so that the total number of particles is conserved.

A particle configuration $C$ can be specified by assigning $1$'s and $0$'s to sites with particles and holes, respectively. Alternatively, one can track the headway $d_i$ which is the number of holes in front of particle $i$. In this way  $d_1,\dots,d_N$ specifies the configuration and the steady-state master equation for $P(d_1,\dots,d_N)$ is given by
\begin{equation}
\fl\quad 0=\sum_{i=1}^{N}[P(\dots,d_{i-1}-1,d_{i}+1,\dots)\theta(d_{i-1}-1)-P(\dots,d_{i},d_{i+1},\dots)\theta(d_{i}),
\label{master_headway}
\end{equation}
\noindent where $\theta(n)=0$ for $n<0$ and $\theta(n)=1$ for $n\geq 1$. Notice that (\ref{master_headway}) is also a steady-state master equation of the zero-range process with single-site weight $f(d_i)=1$; these two processes are thus equivalent.  

The solution to (\ref{master_headway}) is a constant and equals 
\begin{equation}
P(d_1,\dots,d_N)=\frac{1}{{L-1 \choose N-1}}.
\end{equation}
\noindent In the context of vehicular traffic one is often interested in the headway distribution $p(d)$, defined here as the marginal distribution of $P(\{d_i\})$. A straightforward calculation for $p(d)$ gives
\begin{equation}
p(d)=\frac{{L-d-2 \choose N-2}}{{L-1 \choose N-1}}\approx\rho(1-\rho)^d,
\end{equation}
\noindent which for large $N$ and $L$ becomes a geometric distribution with mean $(1-\rho)/\rho$ and variance $(1-\rho)/\rho^2$. Apart from local fluctuations, one can also look at how far the system as a whole is away from its mean,
\begin{equation}
\sigma=\frac{\sum_{i=1}^{N}(d_i-\langle d\rangle)^2}{N}.
\end{equation}
\noindent Here $\sigma$ is precisely the sample variance of the random variables $d_i$ which are constrained to satisfy $\sum_i d_i = L-N$. Thus the analysis of Section \ref{analysis}  applied here predicts that for some finite $\sigma>\sigma_c$\footnote{The random variables here are discrete and thus $\sigma_c$ may differ from the prediction $2\mu^2=2(1-1/\rho)^2$ for continuous variables.} and large $N$ and $L$ condensation will occur. That is for $\sigma >\sigma_c$ we enter a regime where  typical configurations that create a rare event corresponding to a large value of $\sigma$, will contain one large headway of size $O(\sqrt{L})$ and the rest will have sizes of $O(1)$.


\subsection{Condensation transition in a system of polydisperse hard spheres}

Our next example is a system of hard spheres with variable size, proposed in \cite{EMPT10} to sample polydispersity in a hard-sphere fluid. The model consists of $N$ spheres diffusing on an one-dimensional ring of length $L$ and exchanging volume with hard-core interactions. The volume of $i$-th sphere is denoted with $v_i$ and its diameter is $l_i=v_{i}^{1/p}$. Here $p$ is a parameter that takes the value of $p=1$ for rods, $p=2$ for disks, $p=3$ for spheres; in the limit $p\rightarrow\infty$, the spheres become monodisperse. 

Let us designate by $x_i$ the distance between the left-hand sides of two neighbouring spheres. Due to hard-core interactions and periodic boundary conditions, we have the following constraints:
\begin{equation}
\sum_{i=1}^{N}v_i=V,\quad \sum_{i=1}^{N}x_i=L\quad \textrm{and} \quad x_i\geq l_i=v_{i}^{1/p},\quad i=1,\dots,N.
\label{constraints_spheres}
\end{equation}

\noindent Note that whereas the first two constraints are global, the last constraint is a hard-core local constraint. It is also assumed that $L,V$ and $N$ are large such that the ratios $\rho=N/L$ and $\phi=V/N$ are fixed; notice that $1/\rho\geq\phi^{1/p}$ for $p\leq 1$. 

Under certain conditions for the hopping and volume exchange rates the model admits a factorised steady state of the form
\begin{eqnarray}
P(\{x_i,v_i\})&=&\frac{1}{Z_N(L,V)}\left[\prod_{i=1}^{N}b(v_i)\theta(x_i-v_{i}^{1/p})\right]\delta\left(\sum_{i=1}^{L}x_i-L\right)\nonumber\\
&&\times\delta\left(\sum_{i=1}^{L}v_i-V\right).
\end{eqnarray}
\noindent where $Z_{N}(V,L)$ is the partition function of the microcanonical ensemble and $b(v)$ is a function that enters in the expression for the volume exchange rate.

Rather than working with $Z_{N}(V,L)$ directly, we consider the grand canonical partition function $\tilde{Z}_N(\zeta,\eta)$,
\begin{eqnarray}
\fl\quad\tilde{Z}_N(\zeta,\eta)&=&\int_{0}^{\infty}dL \int_{0}^{\infty}dV Z_{N}(L,V)\rme^{-\zeta L-\eta V}\nonumber\\
\fl\quad &=&\int_{0}^{\infty}dx_1\dots dx_L\int_{0}^{\infty}dv_1\dots dv_L \left[\prod_{i=1}^{N}b(v_i)\theta(x_i-v_{i}^{1/p})\rme^{-\zeta x_i-\eta v_i}\right]\nonumber\\ 
\fl\quad &=&\int_{0}^{\infty}dv_1\dots\int_{0}^{\infty}dv_L\left[\prod_{i=1}^{N}b(v_i)\left(\rme^{-\zeta v_{i}^{1/p}}/\zeta\right)\rme^{-\eta v_i}\right]\nonumber\\
\fl\quad &=&[G(\zeta,\eta)]^{N}.
\end{eqnarray}
\noindent Here in the last step the integrals decouple into a product of functions $G(\zeta,\eta)$ given by
\begin{equation}
G(\zeta,\eta)=\frac{1}{\eta}\int_{0}^{\infty} \rmd v\, b(v)e^{-\zeta v-\eta v^{1/p}},
\end{equation}
\noindent A connection with the microcanonical ensemble is established by enforcing the conservation of $L$ and $V$ on the corresponding averages in the grand canonical ensemble yielding the equations,
\begin{equation}
\phi=-\frac{\partial}{\partial \zeta}\textrm{ln}G(\zeta,\eta),
\label{phi}
\end{equation}
\begin{equation}
\frac{1}{\rho}=-\frac{\partial}{\partial\eta}\textrm{ln}G(\zeta,\eta).
\label{rho}
\end{equation}
\noindent The form of $G(\zeta,\eta)$ is equivalent to that of $g(\zeta,\eta)$ in (\ref{g}) using $f=b$ and with an additional factor $1/\eta$,
\begin{equation}
G(\zeta,\eta)=\frac{g(\zeta,\eta)}{\eta}
\end{equation}
\noindent We can also rewrite (\ref{phi}) and (\ref{rho}) using (\ref{z_q}),
\begin{equation}
\phi=z_1(\zeta,\eta),
\label{phi2}
\end{equation}
\begin{equation}
\frac{1}{\rho}=\frac{1}{\eta}+z_{1/p}(\zeta,\eta).
\label{rho2}
\end{equation}
\noindent For $p<1$, equation (\ref{phi2}) can be solved to give $\zeta$ for any given $\eta\geq 0$, as we showed earlier for (\ref{saddle_mu}). The same is true for (\ref{rho2}) due to an additional term $1/\eta$ that diverges in the limit $\eta\rightarrow 0$; this is contrary to the situation we had in (\ref{saddle_sigma}) where the r.h.s. attained a finite maximum at $\eta=0$. As a consequence, there is no condensation scenario for polydisperse
$p$-spheres with $p\leq 1$. 

On the other hand, for $p>1$ $\zeta$ must not be negative which leads to the maximum allowed value for $z_1(\zeta,\eta)$ when $\zeta=0$. The condensation transition thus takes place for $\phi>\phi_c(\rho)$ where $\phi_c(\rho)$ is given by
\begin{equation}
\phi_c(\rho)=z_{1}(0,\eta_{\rho}(0))
\end{equation}
\noindent and $\eta_{\rho}(0)$ solves the equation (\ref{rho2}) with $\zeta=0$. In the condensed phase a single sphere takes a macroscopic fraction $1-\phi_c/\phi$ of the total volume $V$.  
Note that in the special case $b(v)=1$ the transition point $\sigma_c$ can be calculated exactly \cite{EMPT10} and reads
\begin{equation}
\phi_c(\rho)=\frac{1}{[\rho(p+1)]^p}\frac{\Gamma(2p)}{\Gamma(p)}.
\end{equation}

Finally, we refer to a related system of microdroplets that exhibits similar condensation scenario induced by fixing both the total volume and the total surface of microdroplets, where the the latter is due to a fixed amount of surfactants \cite{CuestaSear02}.


\subsection{Entanglement of a bipartite random pure state}

Another example where the presence of two hard constraints drives the system to
exhibit a condensation transition can be found in the computation of
the distribution of entanglement entropy in a random pure state (for a
short review see \cite{SM_review}).
Consider a bipartite quantum system whose Hilbert space is composed
of two smaller subsystems ${\cal H}_A \otimes {\cal H}_B$. 
Let $L$ and $M$ denote the
dimensions of ${\cal H}_A$ and ${\cal H}_B$ and without any loss of generality,
let $L\le M$. For example, one can think of $A$ as a system of interest and $B$
as a heat bath.
The main question of interest is: if we randomly pick a pure state of the full
system, i.e., a normalised state $|\psi \rangle$ (such that $\langle \psi | \psi \rangle = 1$), 
how much `quantum correlation' (measured by the entanglement entropy)
between the two subsystems $A$ and $B$ is present in this random pure state?

The density matrix operator of the full system in this pure state is simply, 
${\hat \rho}=|\psi\rangle \langle \psi|$ with ${\rm Tr}[\hat \rho]=1$.
One then traces out the degrees of freedom of one of the subsystems (say $B$),
and considers the reduced density matrix of $A$: ${\hat \rho}_A= {\rm Tr}_B {\hat \rho}$.
Evidently, since ${\rm Tr}{\hat \rho}=1$, we also have ${\rm Tr}{\hat \rho}_A=1$.
Then the Renyi entanglement entropy, parametrised by $q>0$ and measuring
the entanglement of the subsystem $A$ with $B$, is defined as
\begin{equation}
S_q = \frac{1}{q-1}\, \ln \Sigma_q\,\,\,\, {\rm where}\,\, \Sigma_q= {\rm Tr}[{\hat \rho}_A^q].
\label{Renyi1}
\end{equation}
The limit $q\to 1$ corresponds to the von Neumann entropy, $S_{q\to 1}=-{\rm Tr}[
{\hat \rho}_A \, \ln {\hat \rho}_A]$. The operator ${\hat \rho}_A$ has $L$ nonnegative 
eigenvalues $\{\lambda_1,\lambda_2,\ldots, \lambda_L\}$ which sum up to unity,
$\sum_{i=1}^L \lambda_i=1$. In terms of the eigenvalues $\{\lambda_i\}$, the Renyi
entropy is then given by
\begin{equation}
S_q= \frac{1}{q-1}\, \ln \Sigma_q\,\,\,\, {\rm with}\,\, \Sigma_q= \sum_{i=1}^L \lambda_i^q.
\label{Renyi2}
\end{equation}

When the pure state $|\psi\rangle$ is picked randomly among all
possible normalisable states of the full system, the eigenvalues $\{\lambda_i\}$ of
${\hat \rho}_A$ also become random variables (but still satisfying the constraint
that their sum is unity). Consequently, the entropy $S_q$, or equivalently,
the quantity $\Sigma_q= \exp[(q-1) S_q]= \sum_{i=1}^L \lambda_i^q$ 
is a random variable and one is interested in the probability
distribution of $S_q$ (or equivalently that of $\Sigma_q$). When the pure state
$|\psi\rangle$ is picked uniformly (according to a uniform Haar measure), it induces
a joint pdf of the eigenvalues that is well known~\cite{Page93} 
\begin{equation}
P_L\left(\{\lambda_i\}\right)= A\, \prod_{i=1}^L \lambda_i^{M-L}\, \prod_{j<k} (\lambda_j-\lambda_k)^2\, 
\delta \left(\sum_{i=1}^L \lambda_i-1\right)
\label{jpdf1}
\end{equation}
where $A$ is just the overall normalisation constant and the delta function in the measure
imposes the hard constraint that the trace is unity.  
Consequently the probability density of $\Sigma_q$ is given by
\begin{eqnarray}
P_L\left(\Sigma_q\right)= && A\, \int \prod_{i=1}^L d\lambda_i \lambda_i^{M-L}\, \prod_{j<k} (\lambda_j-\lambda_k)^2\,
\delta \left(\sum_{i=1}^L \lambda_i-1\right)\nonumber \\
&&\times  \delta\left(\sum_{i=1}^l \lambda_i^q-\Sigma_q\right)\, . 
\label{entropy_pdf1}
\end{eqnarray}

Thus, the eigenvalues $\lambda_i$ can be treated like the mass variables $m_i$ as in 
Eq. (\ref{fss})
and formally the computation of the entropy reduces to computing
a multiple integral in the presence of two hard constraints, in a similar fashion to Eq. (\ref{ZMV}).
There is however one important difference between the measure in Eq. (\ref{jpdf1}) and
that of Eq. (\ref{fss}). In Eq. (\ref{fss}), the eigenvalues are {\em non-interacting}
apart from their global constraint on the sum. In contrast, in Eq. (\ref{jpdf1}), the
eigenvalues, apart from the global constraint of having their sum to be unity, also
have explicit pairwise-interaction through the Vandermonde term $\prod_{j<k} (\lambda_j-\lambda_k)^2$.
In spite of this difference, a condensation transition was found to occur in the distribution
of $P_L(\Sigma_q)$ when $\Sigma_q$ exceeds a critical value~\cite{Nadal10,Nadal11}.
In this case, the largest eigenvalue $\lambda_{\rm max}$ becomes much larger
than the rest of the $(N-1)$ eigenvalues. 
For the special case $q=2$, the same transition was also found in Ref.~\cite{Depasquale10}. 
It turns out that due to the presence of pairwise interaction between the eigenvalues,
the speed of the convergence (with size $L$) of the
large deviation probability $P_L(\Sigma_q)$ is however different in this problem~\cite{Nadal10,Nadal11}
compared to the simple noninteracting mass transport models studied in the present paper. Nevertheless, 
the fact that the presence of two hard constraints can drive a condensation transition remains
robust even in presence of pairwise interactions.

\subsection{Simplified model of breathers in Discrete Non-linear Schr\"{o}dinger equation}

It is known that in the discrete non-linear Schr\"{o}dinger equation (DNLSE) localised `breather' solutions are exhibited. Such structures emerge  when  the energy is raised  above some  critical value \cite{Rumpf}. Breathers are essentially non-linear oscillators and are thought to be generated in the solution in order to satisfy the dual constraints of conserved norm of the wavefunction and energy. The statics and dynamics of the breathers has been studied extensively \cite{Rasmussen00,Johansson04,FG08}.

The presence of two constraints in the DNLSE is reminiscent of the problem studied in this work. However the presence of  phase dynamics of the wavefunction makes the DNLSE a more complicated problem. Recently a simplified model, intended to capture  entropic effects and ignoring the phase dynamics was suggested  by Iubini, Politi and Politi \cite{IPP14}. They replaced the deterministic DNLSE with a probabilisitc dynamics that essentially is a realisation of the two constraint problem studied here in the case $p=1/2$.

\section{Conclusion}
\label{conclusion}

In this work we have studied how two global constraints in the form of
linear statistics (\ref{Vdef}) can produce condensation in factorised
stationary states.  Our results show that condensation may occur when
the underlying single site weight $f(m)$ is light tailed in contrast
to the standard condensation involving only one constraint which
requires heavy-tailed $f(m)$.  Unexpectedly, we find that for a
heavy-tailed choice of $f(m)$  the standard condensation may be
suppressed by the second constraint.

We have studied in detail the partition function $Z_L(M,V)$ and the marginal 
distribution $p(m)$. In the condensation regime, the marginal distribution 
displays a bump, which turns out to 
be non-gaussian in $m$ (the bump, in fact, originates from a multivariate 
gaussian due to the interplay between the two statistics).
Moreover, the peak is shifted from the naively expected value of $m$.
We have confirmed our predictions through Monte Carlo simulations performed 
using the algorithm of \cite{IPP14} for the case $p=1/2$ which corresponds 
to a constrained sample variance.

To calculate the marginal distribution we used the equivalence
between factorised steady states and sums of iid random variables. For the 
latter there is a well developed large deviation theory 
which can be used to describe the fluid phase. However, the theory does not apply 
in the condensed regime where the relevant moment-generating function does 
not exist. We find that in this regime the large deviation speed
has a different scaling to  the usual $L$ dependence in the fluid 
phase, see (\ref{Zcond}).

We have derived the critical line and consequent phase diagrams using 
the saddle point method and alternatively by using the G\"{a}rtner-Ellis 
theorem which provides a rigorous confirmation of our results. The 
derivation of our results for the partition function in Section \ref{analysis} 
involved some justifiable approximation that we believe could be made rigorous. 
Our derivation of the marginal distribution on the other hand used an heuristic
argument that the random variables $m_i$ in the background fluid could be 
treated as independent. For the case of a single constraint such a result 
has been proven rigorously \cite{GrosskinskySchutzSpohn03,ArmendarizLoulakis11}.

It is interesting to note that our constraint-driven condensation scenario bears some 
resemblance to condensation previously observed in interacting particle systems with 
two species of particles \cite{EH03,EH04,Grosskinsky08}. There both particle species have 
conserved number of particles, and the dynamics of one species is dependent on the
local distribution of the other species. If the interaction between species 
is attractive, there is a phase in which the condensate of one of the particle species 
(of size $\propto L$) co-exists with a ``weak'' condensate of particles of the other 
species (of size $\propto L^{1/2}$) at the same site, similar to our problem for 
$p=1/2$.

Finally, the large deviation framework allows us to view the problem not
necessarily as one of two constraints but rather as realisation of a
rare fluctuation \cite{MK10}. Recently, there has been increasing interest in
identifying the structure of rare but important fluctuations in
non-equilibrium systems revealing fluctuations that have interesting co-operative 
structure \cite{HarrisRakosSchuetz05,MK10,HurtadoGarrido11,CorberiZannetti13,ZannettiGonnella14}.
Our results add to that context by providing an example of a rare fluctuation 
exhibiting a higher level of organisation and a broken symmetry in the sense that one
random variable dominates.

As an outlook for future work, we mention that other choices for linear
statistic $V_L$ could also give rise to the condensation. Recently,
condensation was observed in joint statistics of sums $\sum_i \lambda_i$ and 
$\sum_i \textrm{ln}\lambda_i$ \cite{CV14}, where $\lambda_i$'s are correlated random 
variables having similar pairwise interaction as in (\ref{jpdf1}). Remarkably, no
condensation is observed when $\lambda_i$'s are non-interacting (apart from their 
global constraints on the sums), which can be easily checked by solving saddle-point 
equations explicitly for the case $f(\lambda)=1$. This provides an interesting case where
two constraints alone are not sufficient to induce condensation, but also an interaction
is needed. Finally, it would also be interesting to look for condensation-like 
phenomena in time-dependent fluctuations rather than in steady states, e.g. in Markov processes 
conditioned on a rare event \cite{CT14}.

%
\ack
We thank S Iubini, M. Marsili, P Politi and H Touchette for helpful discussions. JSN thanks M 
Marohni\'{c} for suggesting the proof of (ii) in \ref{appendix_a}.
JSN and MRE would like to acknowledge funding from EPSRC under grant number EP/J007404/1. 
SNM acknowledges support by ANR grant 2011-BS04-013-01 WALKMAT. MRE and SNM acknowledges
the hospitality of the GGI, Florence during the workshop ``Advances in Nonequilbrium Statistical Mechanics"
(May-June, 2014) where this work was partially completed. 
%
%
\appendix
\section{Analytic properties of $z_{q}(s,\lambda)$ (\ref{z_q})}
\label{appendix_a}

Recall the definition of $z_q(s,\lambda)$ (\ref{z_q}),
\begin{equation}
\label{zq}
z_q(s,\lambda)=\frac{\int_{0}^{\infty}\rmd m\, m^{q} f(m)\rme^{-sm-\lambda m^{1/p}}}{g(s,\lambda)}.
\end{equation}
\noindent Here we prove the following properties of $z_q(s,\lambda)$:
\begin{itemize}
\item[(i)] $z_q(s,\lambda)$ is a decreasing function of $s$ for fixed $\lambda$ and decreases to zero as $s\rightarrow\infty$;
\item[(ii)] $z_q(s,\lambda)\rightarrow\infty$ as $s\rightarrow-\infty$, provided that $f(m)$ is a monotonically decreasing function of $m$;
\item[(ii)] for a given $\mu>0$ and $\lambda$, let $s_\mu(\lambda)$ solve $z_1(s_{\mu}(\lambda),\lambda)=\mu$, then both $s_\mu(\lambda)$ and $z_{1/p}(s_{\mu}(\lambda),\lambda)$ are decreasing functions of $\lambda$.
\end{itemize}

Properties (i) and (ii) ensure that the solution $s_{\mu}(\lambda)$ to $z_1(s,\lambda)$ is unique. Property (iii) shows that the maximum value of $z_{1/p}(s_{\mu}(\lambda),\lambda)$ depends on the range of allowed values for $\lambda$. The fact that $\lambda$ must be non-negative when $f(m)$ belongs to the case (ii) in (\ref{case2}) or to the case (iii) in (\ref{case3}) leads to the condensation when $\sigma>z_{1/p}(s_{\mu}(0),0)$.

\subsection{Proof of (i)}
First we prove that $z_q(s,\lambda)$ is a decreasing function in $s$, for fixed $\lambda$. To this end, we note that $z_q(s,\lambda)$ can be written as $\langle\langle m^q\rangle\rangle$ where the average is taken with respect to $f(m)\exp(-sm-\lambda m^{1/p})/g(s,\lambda)$. Using this notation, the first derivative $\partial z_q/\partial s$ is given by
\begin{equation}
\frac{\partial z_q}{\partial s}=-[\langle\langle m^{q+1}\rangle\rangle-\langle\langle m^q\rangle\rangle\langle\langle m\rangle\rangle].
\end{equation}
\noindent Now, using the Jensen's inequality, 
\begin{equation}
\langle x^r\rangle\leq \langle x^s\rangle^{r/s}, \quad 0<r<s,
\label{holder}
\end{equation}
\noindent for two pairs of $r$ and $s$, $(r,s)=(q,q+1)$ and $(r,s)=(1,q)$, we get for $q\geq 1$
\begin{equation}
\langle x^{q+1}\rangle\geq \langle x^q\rangle\langle x\rangle, \quad q\geq 1.
\label{holder2}
\end{equation}
\noindent From (\ref{holder2}) it follows that $\partial z_q/\partial s<0$, i.e. $z_q(s,\lambda)$ is decreasing in $s$ for fixed $\lambda$.

To show  that $z_q(s,\lambda)\rightarrow 0$ for $s\rightarrow\infty$ let us look at the integral
\begin{equation}
I(s;\phi)=\int_{0}^{\infty}\rmd m \phi(m)e^{-sm}.
\label{integral}
\end{equation}
\noindent This type of integral appears in $z_q$, where $\phi(m)=m^q f(m)\exp(-\lambda m^{1/p})$ in the numerator and $\phi(m)=f(m)\exp(-\lambda m^{1/p})$ in the denominator. For $s\rightarrow\infty$, the integral in (\ref{integral}) will be dominated by the left end point $m=0$. Let us assume that $\phi(m)$ is well-behaved near $m=0$ and can be written as $\phi(m)=m^{\beta}l(m)$, where $\beta>-1$ and $l(m)$ has $n$ derivatives at $m=0$. In that case the Watson's lemma states that 
\begin{equation}
\int_{0}^{\infty}\rmd m \phi(m)e^{-sm}\sim\sum_{k=0}^{n} l^{(n)}(0)\frac{\Gamma(\beta+k+1)}{s^{\beta+k+1}},\quad s\rightarrow\infty.
\label{watson}
\end{equation}
\noindent Applied to our problem, the lemma yields
\begin{equation}
z_q(s,\lambda)\approx \frac{1}{s},\quad s\rightarrow\infty,
\end{equation}
\noindent i.e. $z_q(s,\lambda)\rightarrow 0$ as $s\rightarrow\infty$.

\subsection{Proof of (ii)}

To prove (ii), it is sufficient to prove that $z_1(s,\lambda)\rightarrow\infty$ when $s\rightarrow-\infty$, since by Jensen's inequality $z_q(s,\lambda)\geq [z_1(s,\lambda)]^q$ for $q\geq 1$. For the $q=1$ case, the idea is to find a function $l(s)$ such that $z_1(s,\lambda)\geq l(s)$ for $s<0$ and such that $\lim_{s\rightarrow-\infty}l(s)=\infty$. 

For simplicity, we will assume that $\lambda\geq 0$. Since $f(m)$ is decreasing and $\textrm{exp}(-sm)$ is an increasing function of $m$ for $s<0$, it always holds that 
\begin{eqnarray}
\label{ineq1}
\fl\qquad && \int_{0}^{\infty}\rmd m \enskip m f(m)\rme^{-sm-\lambda m^{1/p}}=\sum_{n=0}^{\infty}\int_{n}^{n+1}\rmd m m f(m)\rme^{-sm-\lambda m^{1/p}}\nonumber\\
\fl\qquad && \geq \sum_{n=0}^{\infty} n f(n+1)\rme^{-sn-\lambda (n+1)^{1/p}}=\rme^{s}\sum_{n=1}^{\infty} (n-1) f(n)\rme^{-sn-\lambda n^{1/p}}
\end{eqnarray}
Similarly,
\begin{eqnarray}
\label{ineq2}
\fl\qquad && \int_{0}^{\infty}\rmd m \enskip f(m)\rme^{-sm-\lambda m^{1/p}}=\sum_{n=0}^{\infty}\int_{n}^{n+1}\rmd m f(m)\rme^{-sm-\lambda m^{1/p}}\nonumber\\
\fl\qquad && \leq \sum_{n=0}^{\infty} f(n)\rme^{-s(n+1)-\lambda n^{1/p}}=\rme^{-s}\sum_{n=0}^{\infty} f(n)\rme^{-sn-\lambda n^{1/p}}
\end{eqnarray}
By combining (\ref{ineq1}) and (\ref{ineq2}) we get
\begin{equation}
\label{l}
z_1(s,\lambda)\geq \rme^{2s}\left(\frac{\sum_{n=1}^{\infty} (n-1) f(n)\rme^{-sn-\lambda n^{1/p}}}{\sum_{n=0}^{\infty} f(n)\rme^{-sn-\lambda n^{1/p}}}\right).
\end{equation}
Next, we will show that the r.h.s. of (\ref{l}) is our sought function $l(s)$. To this end, we split the summations in (\ref{l}) at $n=N(s)$, where $N(s)$ is an increasing function of $-s$, to be chosen later. Let us define a shorter notation $F(n)=f(n)\textrm{exp}(-\lambda n^{1/p})$. Using (\ref{l}) and the fact that all the summands are positive, we can write 
\begin{equation}
\label{l2}
\fl \quad z_1(s,\lambda)\geq \rme^{2s}\left(\frac{\sum_{n=0}^{N(s)} F(n)\rme^{-sn}}{\sum_{n=1}^{N(s)} (n-1)F(n)\rme^{-sn}}+\frac{\sum_{n=N(s)+1}^{\infty} F(n)\rme^{-sn}}{\sum_{n=N(s)+1}^{\infty} (n-1)F(n)\rme^{-sn}}\right)^{-1}.
\end{equation}
For the terms in the parentheses the following inequalities hold
\begin{eqnarray}
&& \rme^{sN(s)}\sum_{n=1}^{N(s)} (n-1)F(n)\rme^{-sn} \geq \sum_{n=0}^{N(s)}F(n)\rme^{s[N(s)-n]}\nonumber\\
&&-F(0)\rme^{sN(s)}-F(1)\rme^{s[N(s)-1]}+[N(s)-2]F(N(s)),
\end{eqnarray}
\begin{equation}
\rme^{sN(s)}\sum_{n=N(s)+1}^{\infty} (n-1)F(n)\rme^{-sn}\geq N(s)\sum_{n=N(s)+1}^{\infty}F(n)\rme^{s[N(s)-n]}.
\end{equation}
We now choose $N(s)$ such that $[N(s)-2]F(N(s))\textrm{exp}(2s)\rightarrow\infty$ when $s\rightarrow-\infty$. For example, we can take 
\begin{equation}
N(s)=2+\left\lfloor{\frac{\rme^{-2s}}{F(N(s))^2}}\right\rfloor,
\end{equation}
which is increasing function of $-s$, as required. In that case both terms in the parentheses in (\ref{l2}), multiplied by $\textrm{exp}(-2s)$, go to $0$ when $s\rightarrow-\infty$ and thus $z_1(s,\lambda)$ diverges.
 
\subsection{Proof of (iii)}
We can prove that $s_\mu(\lambda)$ is decreasing by taking the derivative of $\mu=z_1(s_{\mu}(\lambda),\lambda)$, which after some algebra gives
\begin{equation}
\frac{\rmd}{\rmd\lambda}s_{\mu}(\lambda)=-\frac{\langle\langle m^{1+1/p}\rangle\rangle-\langle\langle m\rangle\rangle\langle\langle m^{1/p}\rangle\rangle}{\langle\langle m^2\rangle\rangle-\langle\langle m\rangle\rangle^2}\leq 0.
\label{s_mu}
\end{equation}
\noindent The inequality in (\ref{s_mu}) easily follows from (\ref{holder2}) using $q=1$. 

Finally, we can show using (\ref{s_mu}) that $\rmd z_{1/p}(s_{\mu}(\lambda),\lambda)/\rmd\lambda$ can be written as
\begin{equation}
\fl\quad\frac{\rmd}{\rmd\lambda}z_{1/p}(s_{\mu}(\lambda),\lambda)=\frac{[\langle\langle m^{1+1/p}\rangle\rangle-\langle\langle m\rangle\rangle\langle\langle m^{1/p}\rangle\rangle]^2}{\langle\langle m^2\rangle\rangle-\langle\langle m\rangle\rangle^2}-[\langle\langle m^{2/p}\rangle\rangle-\langle\langle m^{1/p}\rangle\rangle]^2
\label{z_1p}
\end{equation}
\noindent Using the following substitution, $X=m$ and $Y=m^{1/p}$, (\ref{z_1p}) can be written as
\begin{eqnarray}
\frac{d}{d\lambda}z_{1/p}(s_{\mu}(\lambda),\lambda)&=&-\frac{\langle\langle (X-\langle\langle X\rangle\rangle)^2\rangle\rangle\langle\langle (Y-\langle\langle Y\rangle\rangle)^2\rangle\rangle}{\langle\langle (X-\langle\langle X\rangle\rangle)^2\rangle\rangle}\nonumber\\
&&+\frac{\langle\langle(X-\langle\langle X\rangle\rangle)(Y-\langle\langle Y\rangle\rangle)\rangle\rangle^2}{\langle\langle (X-\langle\langle X\rangle\rangle)^2\rangle\rangle}
\end{eqnarray}

\noindent which is by Cauchy-Schwartz inequality always non-positive.

\appendix
\setcounter{section}{1}
\section{Calculation of the rate function $J(\mu,\sigma)$}
\label{appendix_b}

Here we calculate the rate function $J(\mu,\sigma)$ using standard result from the large deviation theory, the G\"{a}rtner-Ellis theorem. To this end, we start with $L$ random variables conditioned on the value of their sum, $M_L=M$, whose probability density $P(\{m_i\}\vert M_L=M)$ is given by (\ref{fss}). In that context, $Z_L(M,V)$ can be written as
\begin{eqnarray}
Z_L(M,V)&&=Z_L(M)\int_{0}^{\infty}\rmd m_1\dots \rmd m_L P(\{m_i\}\vert M_L=M)\delta(V_L-V)\nonumber\\
&&\equiv Z_L(M)P(V_L=V).\label{ZMV_ld}
\end{eqnarray}
\noindent According to the G\"{a}rtner-Ellis theorem \cite{Ellis95,Touchette09}, the  probability density function $P(V_L=V)$ satisfies the following large deviation principle
\begin{equation}
\label{K}
P(V_L=V)\sim \rme^{-LK(\mu,\sigma)},\quad K(\mu,\sigma)=\mathop{\rm{max}}_{\lambda}\{\sigma\lambda-\kappa(\lambda;\mu)\},
\end{equation}
\noindent where $\sim$ means that $K(\mu,\sigma)=(1/L)\lim_{L\rightarrow\infty}\ln P(V_L=V)$; $\kappa(\lambda;\mu)$ is the scaled cumulant-generating function 
\begin{displaymath}
\kappa(\lambda;\mu)=\mathop{\textrm{lim}}_{L\rightarrow\infty}\frac{1}{L}\textrm{ln}\left\langle \rme^{\lambda V_L}\right\rangle,
\end{displaymath}
\noindent where the average is taken with respect to $P(V_L=V)$. To calculate $\kappa(\lambda;\mu)$, we can write $\langle\textrm{exp}(\lambda V_L)\rangle$ as
\begin{equation}
\left\langle \rme^{\lambda V_L}\right\rangle=\frac{[g_{LD}(0,\lambda)]^L \Lambda_L(M)}{Z_L(M)},
\label{V_mgf}
\end{equation}
\noindent where $g_{LD}(s,\lambda)=g(-s,-\lambda)$ \footnote{Notice that the moment-generating function and Laplace transform (when applied to a probability density function) have opposite sign conventions. To make our calculations easier to follow, it proves easier to keep both conventions by introducing $g_{LD}(s,\lambda)=g(-s,-\lambda)$.} and $\Lambda_L(M)$ reads
\begin{equation}
\Lambda_L(M)=\int_{0}^{\infty}\rmd m_1\dots \rmd m_L\prod_{i=1}^{L}\left[\frac{f(m_i)\rme^{\lambda m_{i}^{1/p}}}{g_{LD}(0,\lambda)}\right]\delta(M_L-M).
\label{lambda}
\end{equation}
\noindent The expression in (\ref{lambda}) has a simple interpretation: it is the probability density for the sum of iid random variables with common  distribution  $f(m)\exp(\lambda m^{1/p})/g_{LD}(0,\lambda)$. We can now use the fact that the moment-generating function of this distribution is given by $[g_{LD}(s,\lambda)/g_{LD}(0,\lambda)]^L$ and then apply the G\"artner-Ellis theorem, which yields
\begin{equation}
\Lambda_{L}(M)\sim \exp\{-L[\textrm{ln}g_{LD}(0,\lambda)+\mathop{\textrm{max}}_{s}\{\mu s-\textrm{ln}g_{LD}(s,\lambda)\}]\}.
\label{lambda_ldp}
\end{equation}
\noindent Similarly, the large-$L$ behaviour of $Z_L(M)$ is given by
\begin{equation}
Z_L(M)\sim \rme^{-L I(\mu)},\quad I(\mu)=\mathop{\textrm{max}}_{s}\{\mu s-\textrm{ln}g_{LD}(s,0)\}.
\label{ZM_ldp}
\end{equation}
\noindent Inserting (\ref{lambda_ldp}) and (\ref{ZM_ldp}) in (\ref{V_mgf}) gives the following expression for the scaled cumulant-generating function $\kappa(\lambda;\mu)$
\begin{equation}
\label{kappa_2}
\kappa(\lambda;\mu)=I(\mu)-\mathop{\textrm{max}}_{s}\{\mu s-\textrm{ln}g_{LD}(s,\lambda)\}.
\end{equation}
\noindent Inserting (\ref{kappa_2}) into (\ref{K}) and then using (\ref{ZMV_ld}) yields the sought rate function $J(\mu,\sigma)$,
\begin{equation}
\label{J_2}
J(\mu,\sigma)=\mathop{\textrm{max}}_{\lambda}\{\sigma\lambda+\mathop{\textrm{max}}_{s}\{\mu s-\textrm{ln}g_{LD}(s,\lambda)\}\}.
\end{equation}
For a given $\lambda$, let us denote by $s_\mu(\lambda)$ the $s$ that maximises $\mu s-\textrm{ln}g_{LD}(s,\lambda)$. Since the G\"{a}rtner-Ellis theorem assumes that $\textrm{ln}g_{LD}(s,\lambda)$ exists and is differentiable, $s_\mu(\lambda)$ must solve the following equation,
\begin{equation}
\label{saddle_mu_2}
\mu=\frac{\partial }{\partial s}\textrm{ln}g_{LD}(s,\lambda).
\end{equation}
\noindent It is also straightforward to show that $s_\mu(\lambda)$ indeed maximises $\mu s-\textrm{ln}g_{LD}(s,\lambda)$, since
\begin{equation*}
\frac{\partial^2}{\partial s^2}\left[\mu s-\textrm{ln}g_{LD}(s,\lambda)\right]=-\left[\langle\langle m^2\rangle\rangle-\langle\langle m\rangle\rangle^2\right]<0.
\end{equation*}
\noindent Here $\langle\langle\dots\rangle\rangle$ denotes averaging with respect to $f(m)\exp(sm+\lambda m^{1/p})/g_{LD}(s,\lambda)$; the expression in square brackets is variance and is thus always positive.

Similarly, we can show that the value  $\lambda^*$ that maximises $\sigma\lambda+\mu s_\mu(\lambda)-\textrm{ln}g_{LD}(s_\mu(\lambda),\lambda)$ solves the equation
\begin{equation}
\label{saddle_sigma_2}
\sigma=\left.\frac{\partial }{\partial \lambda}\textrm{ln}g_{LD}(s,\lambda)\right\vert_{s=s^*=s_\mu(\lambda^*),\lambda=\lambda^*}.
\end{equation}
\noindent From here we can switch back to $s\rightarrow -s$, $\lambda\rightarrow-\lambda$ and $g(s,\lambda)=g_{LD}(-s,-\lambda)$ and recover (\ref{saddle_2}) and (\ref{J_1}); this completes our derivation for large-$L$ behaviour in the fluid phase of the partition function $Z_L(M,V)$, using large deviation theory. 

\appendix
\setcounter{section}{2}
\section{Bivariate central limit theorem}
\label{appendix_c}

Consider two sets of random variables $\{x_{i}^{(1)}\}$ and $\{x_{i}^{(2)}\}$ for $i=1,\dots,L$. It is assumed that $x_{1}^{(1)},\dots,x_{L}^{(1)}$ are identically distributed and mutually independent and the same is assumed for $x_{1}^{(2)},\dots,x_{L}^{(2)}$. Let us define a random vector $\mathbf{x}_i$,
\begin{equation}
\mathbf{x}_i=\left(\begin{array}{c}
x_{i}^{(1)} \\
x_{i}^{(2)}\end{array}\right),\quad i=1,\dots,L,
\end{equation}
\noindent and let $\mathbf{e}$ denote its mean, $\mathbf{e}=\langle\mathbf{x}_i\rangle$. Bivariate central limit theorem states that the sample average $\sum_{i=1}^{L}(\mathbf{x_i}-\mathbf{e})/L$, multiplied by $\sqrt{L}$, converges in distribution to a bivariate Gaussian distribution $\mathcal{N}(\mathbf{x})$ with zero mean and covariance matrix $\mathbf{\Sigma}$,
\begin{equation}
\mathcal{N}(\mathbf{x})=\frac{1}{\sqrt{(2\pi)^2\vert\mathbf{\Sigma}\vert}}\rme^{-\frac{1}{2}\mathbf{x}^T{\mathbf{\Sigma}}^{-1}\mathbf{x}}.
\end{equation}
\begin{equation}
\Sigma_{kl}=\langle (x^{(k)}-e_k)(x^{(l)}-e_l)\rangle, \quad k,l\in \{1,2\}.
\end{equation}


\end{document}